%% file: main.tex
\documentclass[10pt,twocolumn,letterpaper]{article}

\usepackage{iccv}              

\input{tex/preamble}

\definecolor{iccvblue}{rgb}{0.21,0.49,0.74}
\usepackage[pagebackref,breaklinks,colorlinks,allcolors=iccvblue]{hyperref}

\newcommand\blfootnote[1]{%
  \begingroup
  \renewcommand\thefootnote{}\footnote{#1}%
  \addtocounter{footnote}{-1}%
  \endgroup
}

\def\paperID{*}
\def\confName{ICCV}
\def\confYear{2025}

\title{Seeing What You Say: Expressive Image Generation from Speech}

\author{Jiyoung Lee$^{1,\dagger}$\thanks{Corresponding author} \quad\quad Song Park$^{\dagger}$ \quad\quad Sanghyuk Chun$^{2,\dagger}$ \quad\quad Soo-Whan Chung$^3$ \vspace{.8em} \\ 
$^1$Ewha Womans University \quad $^2$Princeton University \quad $^3$NAVER CLOUD \vspace{.3em} \\ 
\small{\url{http://mmai.ewha.ac.kr/voxstudio}}
}

\begin{document}

\twocolumn[{
\maketitle
\renewcommand\twocolumn[1][]{#1}
\maketitle
\begin{center}
    \centering
    \vspace*{-1em}
    \includegraphics[width=.9\linewidth]{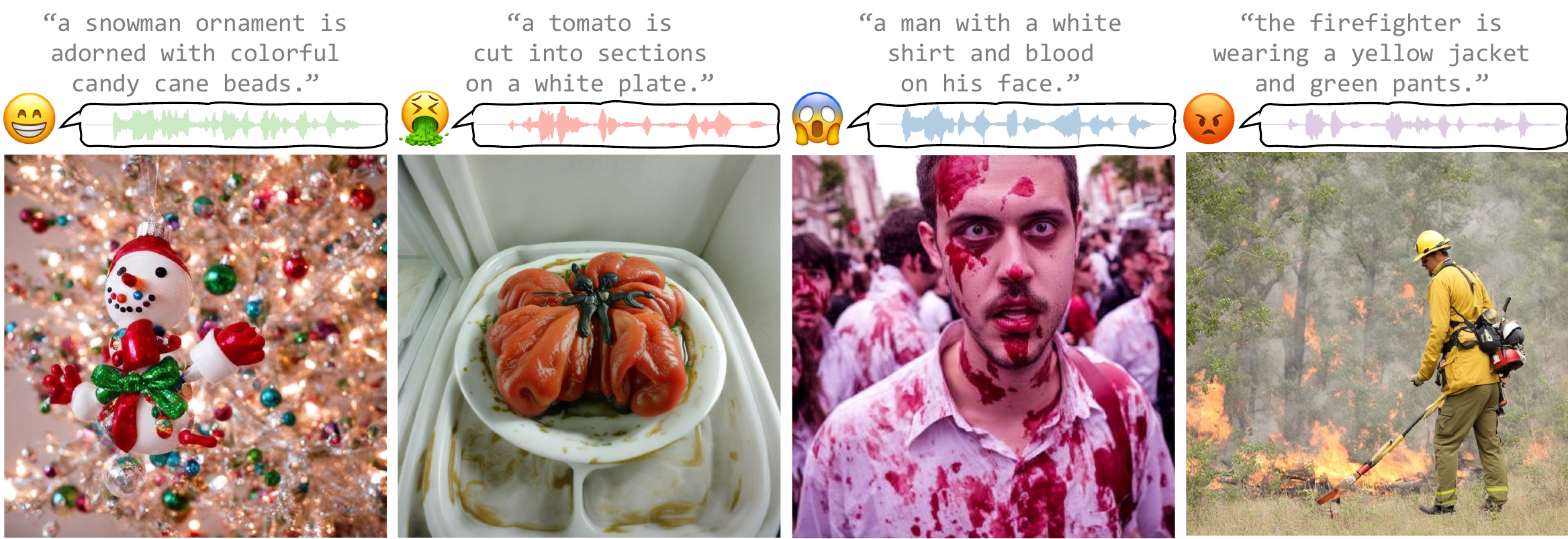}
    \captionsetup{hypcap=false}
    \vspace*{-.5em}
    \captionof{figure}{Generated images by \ours from spoken descriptions.}
\label{fig:teaser}
\end{center}
}]

\input{tex/0_abstract}
\input{tex/1_intro}
\input{tex/2_relatedwork}
\input{tex/3_method}

\input{tex/4_benchmark}
\input{tex/5_experiments}
\input{tex/6_conclusion}

{
    \small
    \bibliographystyle{ieeenat_fullname}
    \bibliography{egbib}
}

\clearpage
\appendix
\setcounter{table}{0}
\setcounter{figure}{0}
\renewcommand\thetable{\thesection\arabic{table}}
\renewcommand\thefigure{\thesection\arabic{figure}}
\input{supp_tex/a_benchmark}
\input{supp_tex/b_result}

\end{document}

%% file: tex/preamble.tex
\usepackage[most]{tcolorbox}

\usepackage{booktabs}
\usepackage{makecell}
\usepackage{amsmath}

\usepackage{tabularx}
\usepackage{pifont} 
\usepackage{amsfonts}
\usepackage{multirow}
\usepackage{graphicx}
\usepackage{caption}
\usepackage{subcaption}
\usepackage[export]{adjustbox}

\definecolor{turquoise}{cmyk}{0.65,0,0.1,0.3}
\definecolor{purple}{rgb}{0.65,0,0.65}
\definecolor{dark_green}{rgb}{0, 0.5, 0}
\definecolor{orange}{rgb}{0.8, 0.6, 0.2}
\definecolor{red}{rgb}{0.8, 0.2, 0.2}
\definecolor{darkred}{rgb}{0.6, 0.1, 0.05}
\definecolor{blueish}{rgb}{0.0, 0.3, .6}
\definecolor{light_gray}{rgb}{0.7, 0.7, .7}
\definecolor{pink}{rgb}{1, 0, 1}
\definecolor{greyblue}{rgb}{0.25, 0.25, 1}

\newcommand{\cmark}{\color{blueish}{\ding{51}}}
\newcommand{\xmark}{\color{light_gray}{\ding{55}}}

\newcommand{\tparagraph}[1]{\noindent\textbf{#1}}

\usepackage{xspace}
\newcommand{\ours}{\texttt{VoxStudio}\xspace}

\newcommand{\oursbench}{VoxEmoset\xspace}
\newcommand{\mapper}{SIB\xspace}

\newcommand{\Appendix}{Appendix\xspace}
\newcommand{\flickr}{Flickr8kAudio\xspace}
\newcommand{\scoco}{SpokenCOCO\xspace}
\usepackage[titletoc,toc]{appendix}

%% file: tex/0_abstract.tex
\begin{abstract}
This paper proposes \ours, the first unified and end-to-end speech-to-image model that generates expressive images directly from spoken descriptions by jointly aligning linguistic and paralinguistic information.
At its core is a speech information bottleneck (SIB) module, which compresses raw speech into compact semantic tokens, preserving prosody and emotional nuance. 
By operating directly on these tokens, \ours eliminates the need for an additional speech-to-text system, which often ignores the hidden details beyond text, \eg, tone or emotion.
We also release \oursbench, a large‐scale paired emotional speech–image dataset built via an advanced TTS engine to affordably generate richly expressive utterances.
Comprehensive experiments on the SpokenCOCO, Flickr8kAudio, and \oursbench benchmarks demonstrate the feasibility of our method and highlight key challenges, including emotional consistency and linguistic ambiguity, paving the way for future research.
\blfootnote{
\hspace{-2em} $^{\dagger}$ Partly work done in NAVER AI Lab.
$^{*}$ Corresponding author.}
\vspace{-2em}
\end{abstract}

%% file: tex/1_intro.tex
\section{Introduction}
Humans naturally “imagine” vivid mental images when listening to speech, which conveys not only semantics but also emotion, tone, and intent. 
Speech-to-image (S2I) generation taps this rich, multimodal expressiveness to produce visuals that are more nuanced and emotionally resonant than those driven by text alone. By translating spoken descriptions directly into images, S2I can unlock applications in accessibility, creative media, and voice‐driven interfaces—treating speech as a first‐class modality for content creation rather than a mere precursor to text.

Recent advances in text-to-image (T2I) generation have demonstrated remarkable progress, but they struggle to utilize the innate expressiveness and accessibility of speech. 
Most cascaded framework—where an utterance is first transcribed into text or textual feature and then used as input for T2I models, as shown in \cref{fig:cascade} (a, b)—encounters several significant challenges.
First, speech-to-text (\ie, ASR) transcription is limited to capturing prosody and speaker intention. 
However, transcription errors propagate into the image generative model, then degrade visual quality.
Second, this sequential approach inherently decouples speech and image generation, making it difficult to transfer crucial prosodic and temporal cues—such as speaking rate, pitch variation, and emotional style—that can influence the mood, color palette, or overall aesthetic of the generated image.
Relying on intermediate text also excludes languages without written forms~\cite{tvlt}.
Even for languages that do have a writing system, coverage for the cascaded approach remains far from comprehensive: there are over 7,100 languages worldwide~\cite{ethnologue}, \eg Google API covers only 125\footnote{https://cloud.google.com/speech-to-text}.
Finally, the cascaded system limits the inference speed and requires a higher cost than our unified system, as shown in the table of \cref{fig:cascade}.
These limitations underscore the necessity of an end-to-end approach that directly maps raw speech to images, enabling a more seamless and expressive integration of modalities.

\input{fig_tex/cascade}
However, incorporating speech input directly into a pre-trained T2I model poses distinct obstacles rooted in the nature of the two modalities.
Speech is a continuous, high-dimensional signal rich in temporal dynamics and spectral detail, whereas T2I models are designed to process compact sequences of token embeddings. 
Bridging this gap requires effective speech representations that can capture both semantic and paralinguistic cues - yet remain mappable to their latent space.
This alignment is complicated by differing tokenization schemes, variable sequence lengths, and unique contextual subtleties inherent to spoken language.

We propose \ours, a novel speech-to-image model that bridges the rich information in speech with the image modality space, enabling more diverse and expressive visual representations.
Building upon T2I models, our framework is suitable for the unique characteristics of speech, which differ from text:
(1) Speech generally contains longer and more variable sequences than text, leading to uneven information density across embeddings.
(2) Speech signals vary significantly depending on speaker identity, recording environment, and emotional state, affecting articulation and duration, even for the same content.
To address these challenges, we introduce a speech information bottleneck (SIB) that efficiently aligns cross-modal latent spaces while preserving key speech features.
Our \mapper encodes compressed conditional features that guide the image generation process.
Through extensive experiments, we establish an effective speech-based guidance for image generation by identifying the optimal combination of speech encoder, \mapper, and image generator.

Our contributions can be summarized as follows:
\begin{itemize}
\item \ours is a unified image generation model with expressive utterance, where both linguistic and paralinguistic cues are compactly captured via the SIB module.
\item We introduce \textbf{VoxEmoset}, an automatically (and efficiently) synthesized dataset of 247k emotional spoken descriptions for sentiment images. VoxEmoset is used for both training and evaluation of S2I.
\item We evaluate \ours in various S2I benchmarks, including SpokenCOCO~\cite{scoco}, Flickr8kAudio~\cite{spokenflickr} and \oursbench, and demonstrate its superiority and high fidelity over baselines.
\end{itemize}

%% file: fig_tex/cascade.tex
\begin{figure}[t]
\centering
\begin{tabularx}{\linewidth}{@{}X @{}}
\includegraphics[width=.45\textwidth, valign=c]{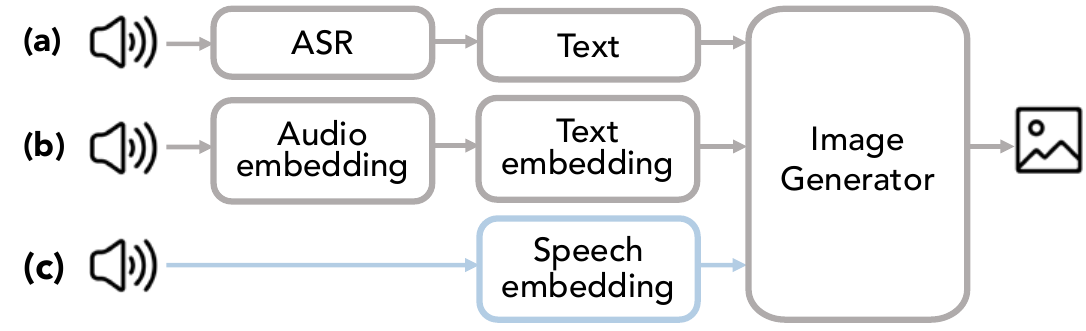}
\hfill \\ 
\vspace{1pt}
\centering
\input{table/cascade.tex}
\end{tabularx}
    \vspace{-5pt}
    \caption{{
    (a) The cascaded system consisting of ASR and T2I, and (b) audio-to-text feature mapping-based methods~\cite{soundbrush,speechclip+} limits in cost than (c) \ours (ours).
    The diffusion process is excluded from GFLOPs and time computations. The parameters of the image generator are also excluded from Params.}}
    \label{fig:cascade}
\end{figure}

%% file: table/cascade.tex
\resizebox{0.47\textwidth}{!}{
\small
\begin{tabular}{lccc}
    \toprule
     & (a) Cascaded & (b) Mapping & (c) \ours \\
    \midrule
    Inference time    &  654.3ms & 23.8ms & 22.2ms \\
    GFLOPS      &  4919G & 154G & 128G \\
    \# Params.      &  2.36B & 1.2B & 0.64B \\
    \bottomrule
\end{tabular}
}

%% file: tex/2_relatedwork.tex
\section{Related Work}

\input{fig_tex/framework}
\tparagraph{Conditions for image generation.}
Recently, diffusion-based conditional image generative models have emerged with remarkable performance \cite{unCLIP, sd, glide,sdxl}. Specifically, stable diffusion (SD)~\cite{sd} has shown impressive results in both quality and generalizability.
Given that these models only take text as a condition, they have struggled to reflect individual thoughts and emotions beyond text into compelling images.
Some methods~\cite{weng2023affective, fu2022language} have proposed emotional image generation, pointing out the importance of reacting to the user's sentiment.
However, they relied on explicit linguistic expressions (\eg, \textit{`with a sense of happiness and joy'} in the text prompt) and focused on reflecting emotion in texture and color only~\cite{borth2013sentibank,brosch2010perception}.
Recently, EmoGen~\cite{yang2024emogen} and EmoEdit~\cite{emoedit} argued that emotional contents beyond color and style should be effectively expressed as semantic variations in a generated image.
They learned a more flexible generative model using a large-scale EmoSet dataset~\cite{emoset}, but still required users to explicitly specify emotion prompts.
In contrast, our approach automatically infers these nuances directly from the speaker’s voice.

In contrast, speech naturally encodes nuanced emotion and tone~\cite{tvlt}, offering a more intuitive means for generating emotionally resonant images, yet it remains largely untapped as a conditioning signal.
Recent audiovisual generation methods~\cite{soundbrush, jeong2023power, cheng2025mmaudio, jeong2025read} have been limited to relying only on semantic instances expressed in text, where the other expressions are excluded. 
Moreover, existing approaches~\cite{s2igan,wang2021generating, tmt} for S2I generation have used highly limited datasets~\cite{spokenflickr}, restricting their expressive versatility.
We aim to design a unified and emotion-driven S2I framework as well as to introduce a large-scale dataset for both training and evaluation.

\tparagraph{Relationship between speech and image.}
Speech-image relationships have been widely explored in biometrics~\cite{facetts,nagrani2018seeing,chung2020perfect}, linguistic alignment~\cite{speechclip+,scoco,tmt}, and phonetic articulation~\cite{wen2021seeking,syncnet,pm}. 
These studies provide valuable insights into how speech and vision interact in different contexts.
Also, image-speech retrieval~\cite{speechclip+,spokenflickr} has explored the alignment between spoken descriptions and images, highlighting the importance of understanding the semantics in both modalities.
Despite these advances, most of the existing studies focus on isolated characteristics between modalities.
By moving beyond traditional mappings, our work aims to bridge the gap by simultaneously leveraging natural semantic correspondences, \ie, both linguistic and paralinguistic information, between speech and vision.

%% file: fig_tex/framework.tex
\begin{figure*}[t]
    \centering
    \includegraphics[width=.9\linewidth]{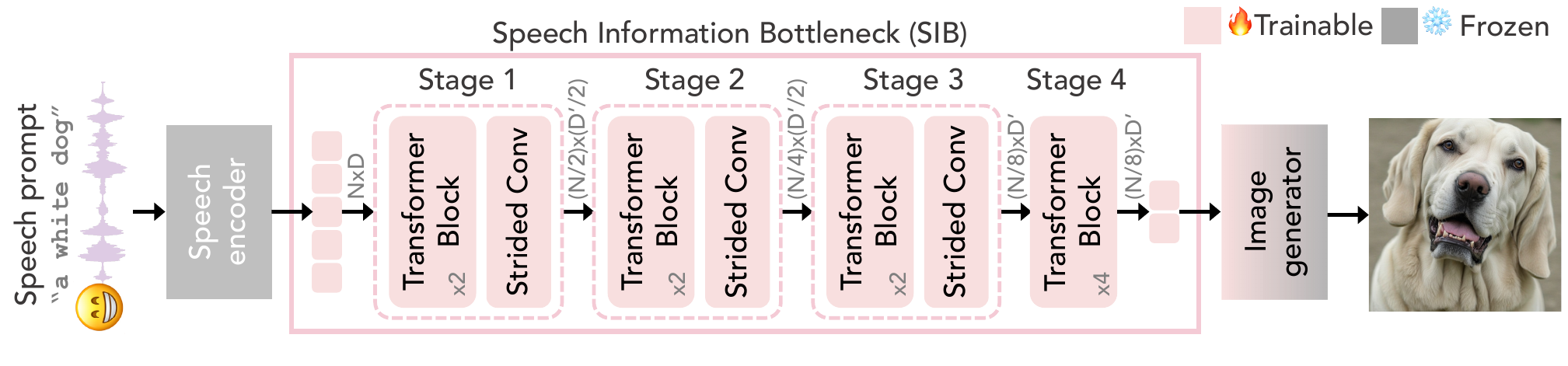}
    \vspace{-1em}
    \caption{\small{\ours encodes an utterance to generate an expressive image. SIB compresses speech embedding into compact semantic tokens to condition the image generator;  training the generator is an optional choice. Trainable parts are optimized by the diffusion loss.}}
    \label{fig:framework}
\end{figure*}

%% file: tex/3_method.tex
\section{\ours}

\cref{fig:framework} shows the overall framework of \ours, consisting of (1) a pretrained speech encoder, (2) \mapper to reduce the computational cost and effectively connect heterogeneous two modalities, and (3) an image generator to synthesize an image from the compressed speech representations.
Next, we describe each module in detail.

\subsection{Speech embedding}
We consider two pre-trained speech encoders, SONAR~\cite{sonar} and Whisper large-v3~\cite{whisper}, to deploy comprehensive speech features considering both linguistic and paralinguistic information.
Briefly reviewed, SONAR~\cite{sonar} has learned global semantic alignment between speech and text, enabling it to encode meaning beyond phonetic content.
We remove its last aggregation layer to ensure that the speech embeddings retain both linguistic and paralinguistic information.
We also test Whisper~\cite{whisper}, a widely used speech recognition model. Whisper is known to be capable of capturing paralinguistic information, such as emotion and speaker identity~\cite{goron2024improving,whispersv}.
Formally, given an input utterance $X$, we obtain a speech embedding $s \in \mathrm{R}^{N \times D}$, where $N$ depends on the length of the speech and models, and $D$ is the channel dimension of the final output layer.
We note that our work does not employ ASR system or text encoder to explicitly map the speech into text.

\subsection{Speech information bottleneck (SIB)}
Although speech embeddings contain rich representations, they are excessively long and lead to a lower information density in each speech token compared to text (\eg, Whisper encodes a maximum of 1500 tokens for 30-seconds long speech, while CLIP text encoder limited to 77 tokens).
This low density makes direct usage challenging to condition the image generator.
To solve this problem, we design a Transformer-based speech information bottleneck (SIB) module.
SIB compacts semantics in speech embeddings, motivated by previous works~\cite{pit, verma2021audio} applied to individual image and audio encoders.
As shown in \cref{fig:framework}, SIB reduces the number of embeddings with a strided convolution layer after a Transformer block along the time axis. 
Based on our findings, a pooling ratio of 8 provides the optimal balance, allowing us to maximize the information retention of speech features.
As a result, the initial embedding $s$ is processed into a compressed speech condition $c = f_{\psi}(s)$, where $c \in \mathrm{R}^{M \times D'}$, $M = N/8$ and $D'$ is the input channel of the cross-attention block in the image generator.
Those compressed representations improve the efficiency of the S2I process while preserving both linguistic and emotional expressiveness.

\subsection{Image generator}

The image generator is based on the latent diffusion model \cite{sd}.
The speech condition $c$, compressed through SIB, is fed into the generator as a guidance of the synthesis process. 
Specifically, the speech embeddings are injected into the UNet through cross-attention layers to condition the image synthesis.
This conditioning allows the model to incorporate the emotional, semantic content of speech into the generation process.
The image generator and SIB are optimized with the diffusion loss~\cite{sd}.
Given that we do not design a specialized loss function compared to previous works such as contrastive learning in~\cite{speechclip+}, AR modeling in~\cite{tmt}, our simple training framework ensures versatile connections for various image generators.
In inference, the denoised latent is decoded into the image through the decoder~\cite{vae}.
Given that we do not design a specialized loss function compared to previous works, such as contrastive learning in~\cite{speechclip+}, AR modeling in~\cite{tmt}, 
our simple framework ensures versatile connections for various image generators.

%% file: tex/4_benchmark.tex
\input{table/benchmark}
\input{fig_tex/scoco_voxemoset}

\section{\oursbench Benchmark}
Our \ours is to generate an image with a corresponding spoken description, even for emotional expression.
However, prior datasets~\cite{spoken-place,scoco} overlooked paralinguistic features in speech, and also required significant costs for human recordings.
Our benchmark uniquely leverages synthesized speech, enabling the natural and cost-effective creation of a large-scale dataset.
Specifically, \oursbench leverages semantic knowledge and the generative powers of pre-trained multimodal LLMs and diffusion models to generate diverse synthetic data samples.
First, a multimodal LLM~\cite{llava} generates corresponding captions that are factual descriptions of a given emotional image based on explaining environments or objects.
Then, a TTS model~\cite{f5tts} generates emotional speech samples from text captions, using emotional voice samples from other datasets as references.
Consequently, we efficiently and cheaply generate large-scale emotional utterances along with text captions, as shown in \cref{fig:samples}.

\subsection{Image collection}
Our benchmark uses images in EmoSet~\cite{emoset}, the large-scale visual emotion dataset annotated with Mikels model~\cite{mikels2005emotional}.
We use the partial of 118k subset labeled by humans and machines, including six categories: amusement, excitement, anger, disgust, fear, and sadness. 
In line with \cite{emobox, ekman, schuller2018speech}, we group amusement and excitement into a single emotion category, `enjoyment', because these two categories are difficult to distinguish solely through voice expression.
On the other hand, we exclude `awe' and `contentment' emotion categories which are hard to express in voice.
The final number of images in \oursbench is shown in \cref{tab:benchmark}.

\subsection{Image caption generation}
While EmoSet categorized emotion classes, there is no sentence-level description for visual scenes.
We generate captions using the instruction prompt in \cref{app:benchmark}, restricting immediate emotional expressions while focusing on factual descriptions.
LLaVA-OneVision~\cite{llava-onevision}, using SigLIP~\cite{siglip} as an image encoder and Qwen-2~\cite{qwen2} as LLM, generates three different captions for each image to prevent the model from simply generating emotionally biased captions, \eg, `\textit{a person is happy.}', `\textit{disgusting rotten egg in the plate.}'.
The word count distribution of our 247k generated captions closely matches that of existing benchmarks, indicating that they were carefully crafted to resemble real-world datasets~\cite{scoco,spokenflickr} (see \cref{app:word_dist} in \Appendix).

\subsection{Speech prompt generation}
Emotional utterances are generated by a text-to-speech (TTS) system that can synthesize the speech with emotional attributes.
This strategy eliminates the dependency on skilled voice actors or noisy crowdsourcing.
Through empirical comparison of recent TTS models based on diffusion, autoregressive, and non-autoregressive architectures, F5-TTS~\cite{f5tts} demonstrates remarkable quality in both linguistic and emotional expression.

Specifically, to build a diverse range of emotional voice references for TTS, emotional speech data was collected from multiple datasets, including CREMA-D~\cite{crema-d}, MEAD~\cite{mead}, and RAVDESS~\cite{ravdess}.
These datasets contain English-spoken utterances from a variety of speakers.
Following EmoBox~\cite{emobox}, we split the datasets into training and test sets.
We validate the emotions in the generated speech using Emotion2Vec~\cite{ma2023emotion2vec} to measure emotional intensity, filtering and re-generating inadequate samples. After this process, 247k speech samples are generated.
Further details are provided in \Appendix.

\subsection{Dataset quality}
To objectively assess the quality of generated utterances, we randomly sample 10k utterances from each dataset and measure NMOS~\cite{dnsmos}. For CREMA-D, we use the entire samples.
\cref{tab:benchmark} shows that \oursbench is compatible with existing speech-image datasets such as SpokenCOCO and Flickr8kAudio in terms of speech quality (NMOS) and description quality (CLIPScore).
However, only our benchmark explicitly expresses emotion in speech.
The last two rows in \cref{tab:benchmark} validates that \oursbench guarantees high perceptual fidelity with clear affect, where emotion discriminability (Emo-C) is measured as the emotion classifier's average confidence score.

%% file: table/benchmark.tex
\begin{table*}[t]
\centering
    \small
    \begin{tabular}{l ccc ccccc}
        \toprule
       Benchmark & \# Images & \# Utterances  & ClipScore & Length (s) & Avg. Words & NMOS & Emotion & Emo-C \\
        \midrule
        SpokenCOCO & 123k & 615k   & 30.42 & 4.34 & 10.45 & 2.9616 &  \xmark & - \\
        Flickr8kAudio & 8k & 40k & 31.27 & 4.12 & 10.87 & 2.9689 &  \xmark & - \\
        \midrule
        CREMA-D & \xmark & 7k   & - & 2.54 & 5.26 & 2.0314 &  \cmark & 0.8465\\
        \midrule
        \oursbench (ours) & 82k & 247k   & 30.27 & 4.25 & 11.19 & 2.9683 & \cmark & 0.8998 \\
        \bottomrule
    \end{tabular}
    \vspace{-.2em}
    \caption{{SpokenCOCO~\cite{scoco}, Flickr8kAudio~\cite{spokenflickr}, and our \oursbench contain paired image-utterance data while CREMA-D~\cite{crema-d} 
    contains utterance only. Our \oursbench shows compatible quality for real-world speech in terms of NMOS and emotional confidence (Emo-C).}}\label{tab:benchmark}
\end{table*}

%% file: fig_tex/scoco_voxemoset.tex
\begin{figure*}[!t]
    \centering
    \begin{subfigure}[b]{0.555\linewidth}
        \includegraphics[width=.93\linewidth]{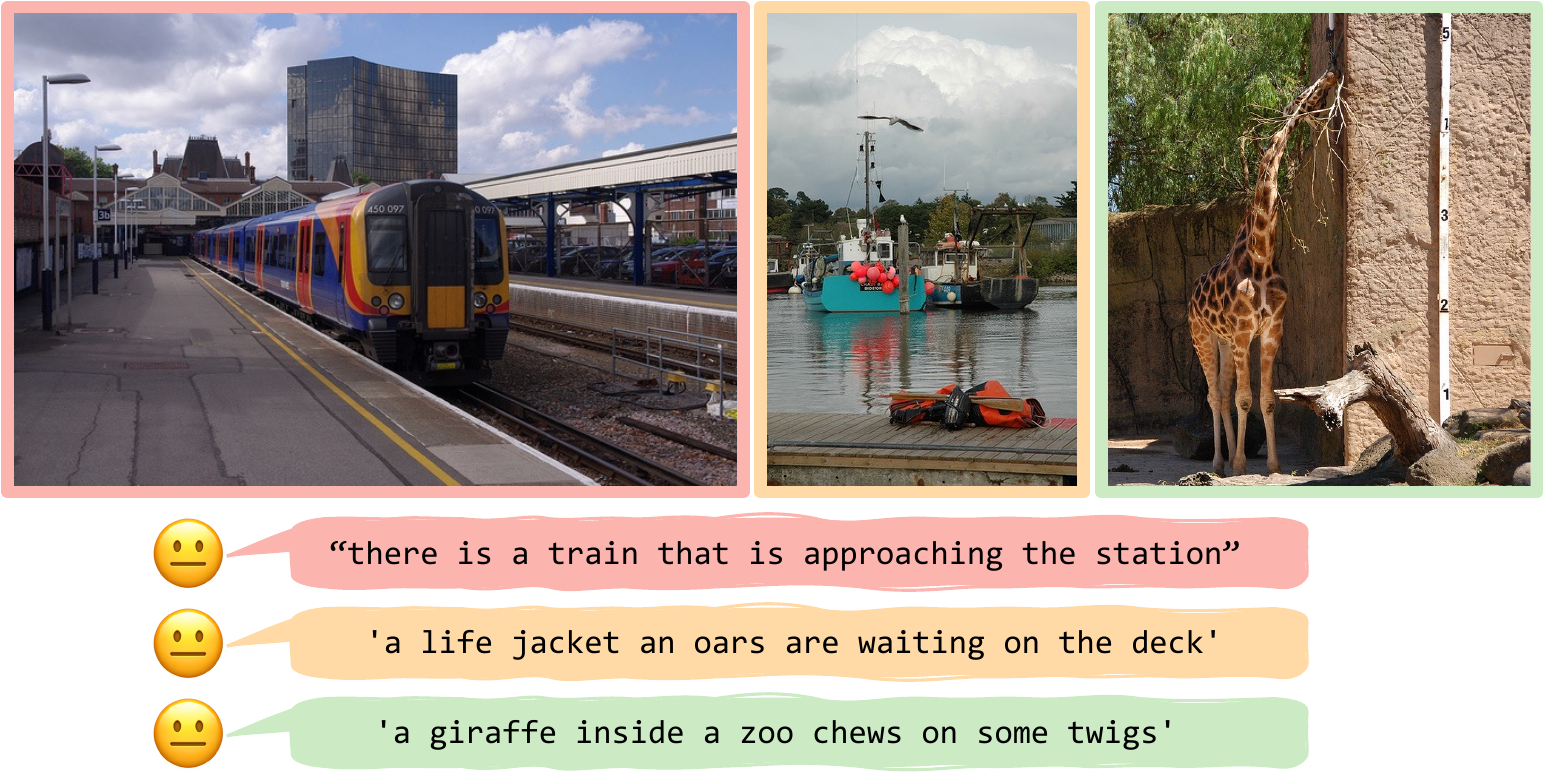}
        \caption{SpokenCOCO \vspace{-1em} }
    \end{subfigure}\hfill
    \begin{subfigure}[b]{0.443\linewidth}
        \includegraphics[width=.93\linewidth]{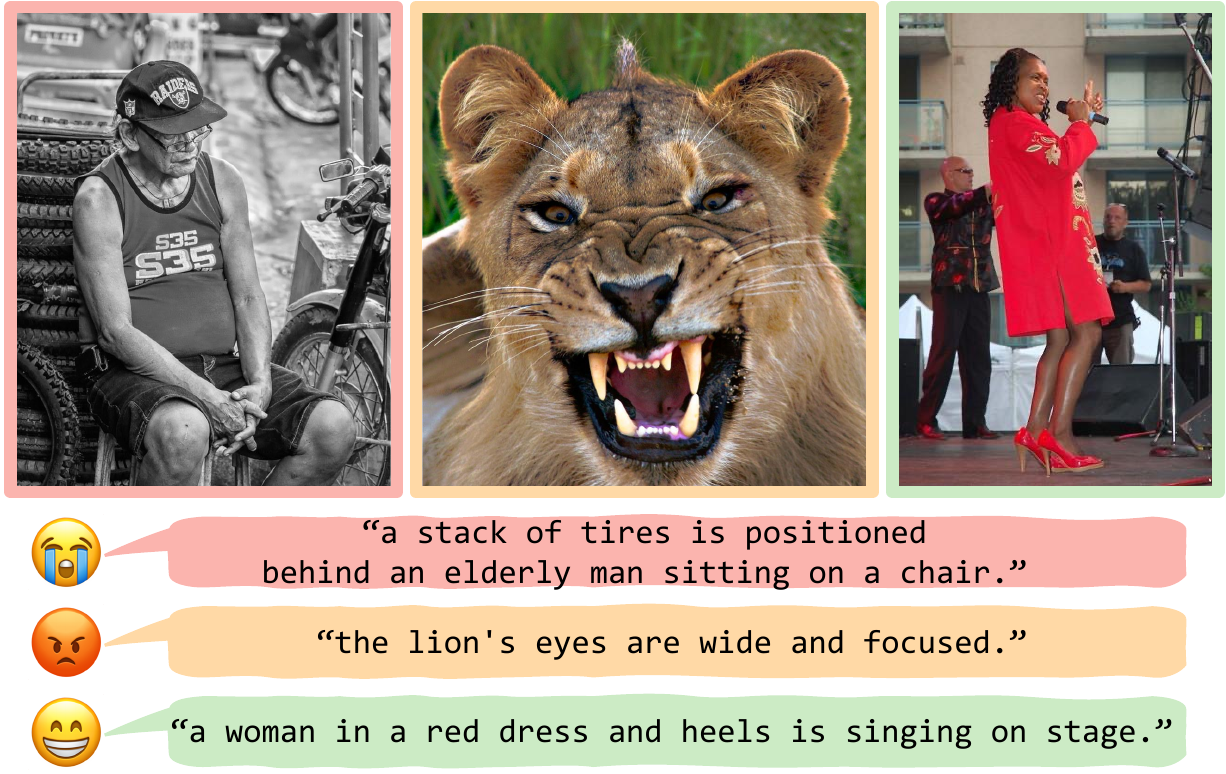}
        \caption{\oursbench \vspace{-1em} }
    \end{subfigure}
    \caption{Examples from \scoco (neutral tone) and \oursbench (expressive tone).}
    \label{fig:samples}
\end{figure*}

%% file: tex/5_experiments.tex
\input{table/main_exp}
\input{fig_tex/sd_ours}

\section{Experiments}

\subsection{Experimental setup}
\tparagraph{Datasets.}
We use SpokenCOCO~\cite{scoco} and \oursbench to train \ours.
\oursbench includes 208k utterances with paired 69k images for training, while SpokenCOCO contains 118k images with 591k utterances.
Flickr8kAudio~\cite{spokenflickr} is used to evaluate zero-shot generalizability.
Each image in SpokenCOCO and Flickr8kAudio has five voice recordings from unskilled annotators, resulting in inherently noisy audio (\eg, the recording may contain background noise, reading speed or volume can vary, and pronunciation may not be as clear as that of skilled voice actors as in \cref{app:limit}).
\oursbench is automatically generated and less prone to recording noise.
We use the Karpathy split~\cite{karpathy2014deep} for \scoco and \flickr.

\tparagraph{Implementation details.}
Training a high-performance image generator requires a vast amount of resources (\eg, SD1.5 requires 6,000 A100 GPU days~\cite{pixart-alpha}). 
We initialize the image generator with a pre-trained SD~\cite{sd} for efficient learning.
We use SONAR as the speech encoder, freezing during the training, in which its last aggregation layer is removed.
We use AdamW \citep{adamw} with the learning rate of $1e$-$6$, the batch size of 128 using 8 V100 GPUs.
FP16 precision is used for all experiments.
The code will be released.

\tparagraph{Evaluation metrics.}
We assess the generation quality using FID~\cite{fid}, while content alignment between speech and generated images is measured with CLIPScore~\cite{clipscore} using text transcriptions.
For SpokenCOCO and \oursbench, random samples of 10k condition prompts, either speech or text, are used for evaluation.
For Flickr8kAudio, we use 5k test prompts for evaluation.
We also report emotion classification accuracy (Emo-A)~\cite{yang2024emogen} on generated images to examine whether the results reflect emotion from prompts.
Note that we measure accuracy only with scores for the 5 emotion categories $-$\textit{`amusement'} and \textit{`excitement'} are classified as the same class$-$ in the trained emotion classifier.

\subsection{Results}
\tparagraph{Results on \scoco and \oursbench.}
\cref{tab:main_exp} shows the comparison of \ours and baselines\footnote{EmoGen was excluded because its pretrained weights are publicly unavailable, and our reimplementation was unable to match its reported performance.} on \scoco and \oursbench.
SD1.5 with the text inputs (\ie, without speech) is shown as a baseline.
Especially, \cref{fig:sd_ours} highlights the stark contrast between text- and speech-based generation. 
While speech conveys emotions even with the same wording, the text-based model inherently ignores these cues and focuses on fact-based generation. 
Even when trained on \oursbench, `SD (finetuning)' struggles to express emotions as semantic content, but speech leads to a richer and intense emotional expression. 
For example, given the prompt `A black trash can is placed against a white wall,' our model detects disgust from spoken nuances and visually emphasizes the unpleasantness of trash, whereas the text-based model remains neutral.

\input{table/spokenflickr}

Furthermore, despite the inherent noise in speech features and our method needs significantly lower latency compared to text-based approaches, the performance gap remains minimal in image quality and text alignment.
Moreover, as shown in the last example in \cref{fig:sd_ours}, the CLIP encoder~\cite{clip} often overlooks information from the latter part of a sentence~\cite{longclip} (\eg, `bright yellow' in the last example).
However, \ours excels in conveying emotions when trained on the same datasets. 
This advocates that speech, as a richer modality for emotional expression, provides a more effective signal to generate emotionally compelling images. 

\input{fig_tex/scoco_flickr}
Remarkably, \ours outperforms SpeechCLIP+ and TMT on \scoco, where \ours does not use \flickr for training.
While TMT additionally used huge synthesized speech data from CC3M~\cite{cc3m} and CC12M~\cite{cc12m} for training, \ours also show comparable results on \oursbench.
This result demonstrates that our diffusion model is a powerful learner for speech-to-expressive image alignment than contrastive learning~\cite{speechclip+} and auto-regressive training~\cite{tmt}.
The qualitative comparison on \scoco shows that SpeechCLIP+ and TMT often ignore keywords in the prompts, while \ours can capture the details, as shown in \cref{fig:scoco_flickr}.

\tparagraph{Results on \flickr.}
\cref{tab:flickr} shows the performance comparison on \flickr.
Here, while TMT and SpeechCLIP+ used \flickr for training, \ours was evaluated in a zero-shot manner.
Surprisingly, \ours outperforms existing methods by large margins.
It shows that end-to-end training in \ours is more robust in aligning the speech-language space.
By contrast, speech features in \ours are more robust to the order or length of the prompt.
Moreover, \oursbench might improve the robustness on generality as shown in \cref{fig:scoco_flickr}.

\input{fig_tex/user_study.tex}
\tparagraph{Human evaluation.}
A user study is conducted to assess how well humans perceive the alignment between speech and image atmosphere. 
26 participants evaluated 25 images to rate how well the emotion conveyed in the image matched the given speech. 
\cref{fig:user_study1} shows that results from \ours are more aligned with the emotion than text-based SD in all categories.
In other words, with an average of 57.09\% preference, the images generated by our \ours were rated as better at expressing emotions.
It highlights the effectiveness of speech prompts for expressive image synthesis.
We also carried out another human evaluation on \scoco across speech-based models.
This experiment is performed on 17 participants who evaluate 10 generated images for each model with a 5-point Likert scale.
As demonstrated in \cref{fig:user_study2}, \ours outperforms existing approaches by generating high-quality images that accurately reflect the nuances of the input prompts.

\input{fig_tex/diff_emo}

\subsection{Discussion}
We note that \ours-FT's performance is basically reported in this section, except \cref{tab:sdxl}.

\input{fig_tex/esd_real}
\input{table/training_data}
\input{table/finetuning}
\tparagraph{Effect of emotion.}
\cref{fig:diff_emo} demonstrates that the same description, when spoken with different emotions, leads to distinct visual outputs by \ours. This highlights \ours's capability to produce emotional nuances beyond linguistic content. 
For instance, a neutral statement spoken in a disgusted tone results in negative visual details (top-left), while an “enjoying” tone generates a more positive scene (top-right).
These findings show that our speech-based approach effectively leverages emotional cues, enabling more expressive and context-rich image generation.

\begin{figure}[t]
    \centering
    \includegraphics[width=\linewidth]{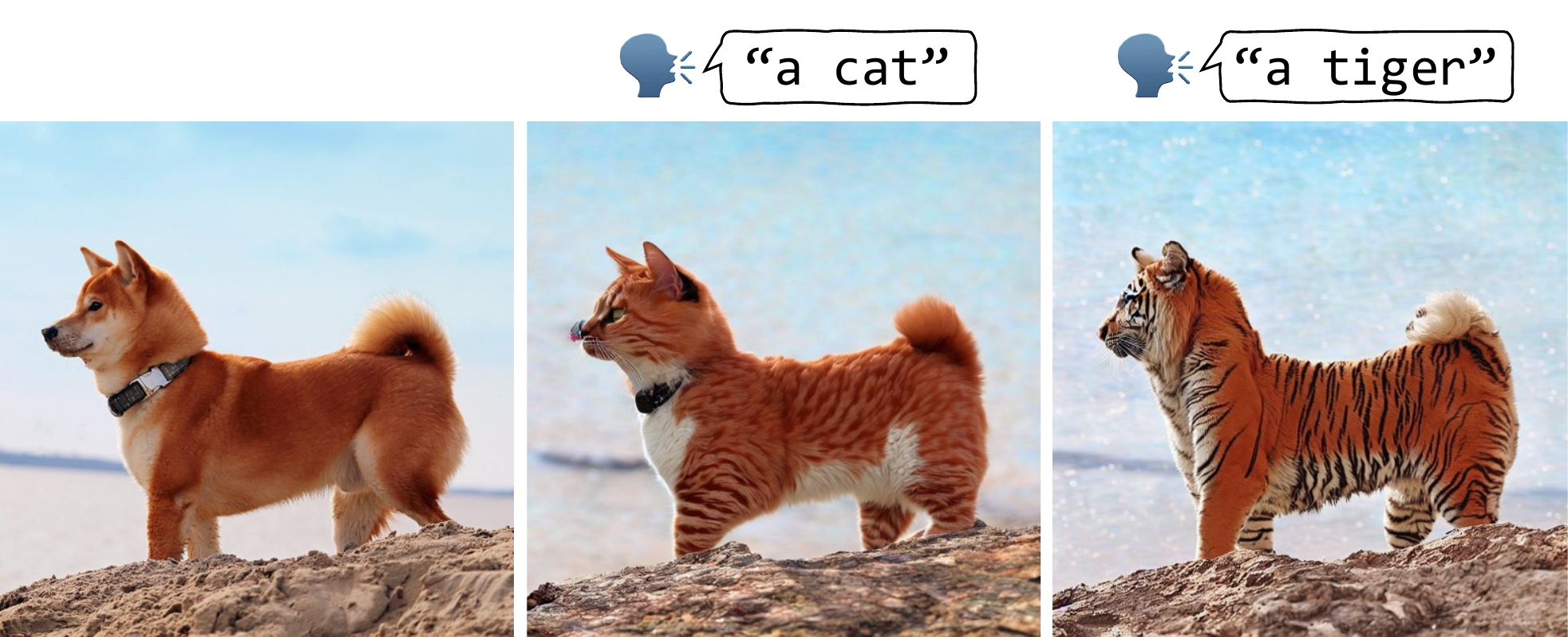}
    \caption{Image editing using speech prompt.}
    \label{fig:i2i}
\end{figure}

\input{table/sdxl_sd15.tex}
\input{table/encoder}
\input{table/mapper}
\tparagraph{Generalization on real speech.}
To evaluate how well our model generalizes, we test on utterances in ESD~\cite{esd} dataset, excluded from our reference samples during speech synthesis.
We visualize generated samples from real speakers' utterances in \cref{fig:realgen}.
Text-based generator is limited to expressing the tone in speech prompt, but \ours successfully expresses atmospheres despite ambiguous words.
It also proves the superiority of \oursbench in that \ours trained on synthesized emotional speech is well generalized in real utterances. 
\cref{fig:i2i} also demonstrates that \ours extends naturally to various applications such as image editing by spoken prompt.

\tparagraph{Training datasets.} \cref{tab:training_data} shows that \oursbench is complementary with the real-world spoken dataset, \scoco, improving both visual fidelity and semantic relevance.

\tparagraph{Training strategies.}
Diffusion training is usually computationally expensive. 
We test different training strategies in \cref{tab:finetuning}: full finetuning, LoRA~\cite{lora}, and freezing the model. 
While finetuning achieves the best performance, LoRA and frozen models show comparable results in CLIPScore and Emo-A.
Additionally, although speech is noisier than text, our method outperforms full finetuning for original SD1.5 (`SD(T2I)-FT' in \cref{tab:finetuning}) in terms of emotional expression, while maintaining the generation quality.

\tparagraph{Scale of the image generator.}
\cref{tab:sdxl} demonstrates the performance of image generators at different scales.
Due to the resource limit, we compare UNet of SD1.5~\cite{sd} and SDXL~\cite{sdxl} as our image generator in a frozen state during the training.
Interestingly, although the small generator achieves a higher CLIPScore, the larger generator excels at displaying emotional nuances.
This finding suggests that larger-scale generators are inherently better at representing content beyond simple text cues.

\tparagraph{Speech embedding.} 
We compare SONAR~\cite{sonar} and Whisper-Large v3~\cite{whisper} encoders as a speech input handler of our method.
Whisper is a widely used ASR model, also known to be capable of capturing paralinguistic information~\cite{goron2024improving,whispersv}.
While SONAR is sentence-level speech-text aligned features, Whisper is trained at the phoneme-level by predicting which words are spoken in a given audio snippet.
This fundamental difference affects how each encoder preserves linguistic content and emotional cues when mapping speech to image descriptions. 
\cref{tab:encoder} demonstrates that text-aligned embeddings (\ie, SONAR) show more robust performance on our task.

\tparagraph{Architecture choices on \mapper.}
We propose \mapper to represent the speech condition compactly to address the issue of low information density of speech tokens.
\cref{tab:mapper} compares its performance against a standard transformer structure.
Our design choice achieves better performance with fewer parameters. 
Gradually reducing the speech token length while simultaneously increasing their density across multiple layers enhances both the linguistic and paralinguistic expressiveness of the speech signal.

\tparagraph{Image editing using speech prompt.}
\ours is built upon the SD architecture, allowing seamless integration with various extensions and applications.
For instance, as shown in \cref{fig:i2i}, the image editing pipeline can be directly applied with speech prompts to modify input images.
Beyond basic editing, our framework can be extended to other tasks built on SD, including personalized generation~\cite{dreambooth, ip-adapter} and multimodal content synthesis~\cite{emma}.
It provides a versatile foundation for future developments in S2I generation.

%% file: table/main_exp.tex
\begin{table*}[t]
    \centering
    \small
    \begin{tabular}{l c c c ccc ccc}
        \toprule
        \multirow{2}{*}[-0.4ex]{Method} & \multirow{2}{*}[-0.4ex]{SD} &  \multirow{2}{*}[-0.4ex]{\makecell{\# training \\ utterances}} & \multirow{2}{*}{Input}   & \multicolumn{2}{c}{(Spoken)COCO} & \multicolumn{3}{c}{\oursbench} \\
        \cmidrule(lr){5-6} \cmidrule(lr){7-9}
        &  & &  
        & FID$\downarrow$ & CLIPScore$\uparrow$ 
        & FID$\downarrow$ & CLIPScore$\uparrow$ & Emo-A$\uparrow$ \\
        \midrule
        T2I & 1.5 & - & Text  & 23.37 & \bf 31.14 & \bf 20.21 & \bf 31.70 & \bf 60.81 \\
        Whisper (ASR) & 1.5 & - & Text  & \bf 22.95 & 31.08 & 20.23 & 31.57 & 60.41 \\
        
        \midrule
        SpeechCLIP+ & 1.5 & 621k & Speech & 28.29 & 25.03 & 33.75 & 21.84 & 37.42 \\
        SpeechCLIP+$\dagger$ & 1.5 & 829k & Speech  & 27.58 & 26.29 & 28.80 & 26.72 & 56.39 \\
        TMT & 2.1 & 15.6M$\ddagger$ & Speech  & \bf 25.48 & 28.26 & 29.48 & 26.08 & 48.54 \\
        \ours & 1.5 & 799k & Speech  & 27.20 & \bf 28.71 & \bf 25.01 & \bf 28.71 & \bf 67.09 \\
        
        \bottomrule
    \end{tabular}
    \vspace{-.7em}
    \caption{{Performance comparison with baselines; SD~\cite{sd}, SpeechCLIP+~\cite{speechclip+} and TMT~\cite{tmt}.  
    `Input' denotes the data type of the input condition for generative models: `T' is text and `S' is speech. SpokenCOCO contains 591k training utterances, Flickr has 30k, and \oursbench includes 208k. All methods were implemented on frozen image generators. $\dagger$: SpeechCLIP+ is finetuned on \oursbench. $\ddagger$: TMT used an additional 15M synthesized speech for training. }}\label{tab:main_exp}
\end{table*}

%% file: fig_tex/sd_ours.tex
\begin{figure}[!t]
    \centering
    \includegraphics[width=\linewidth]{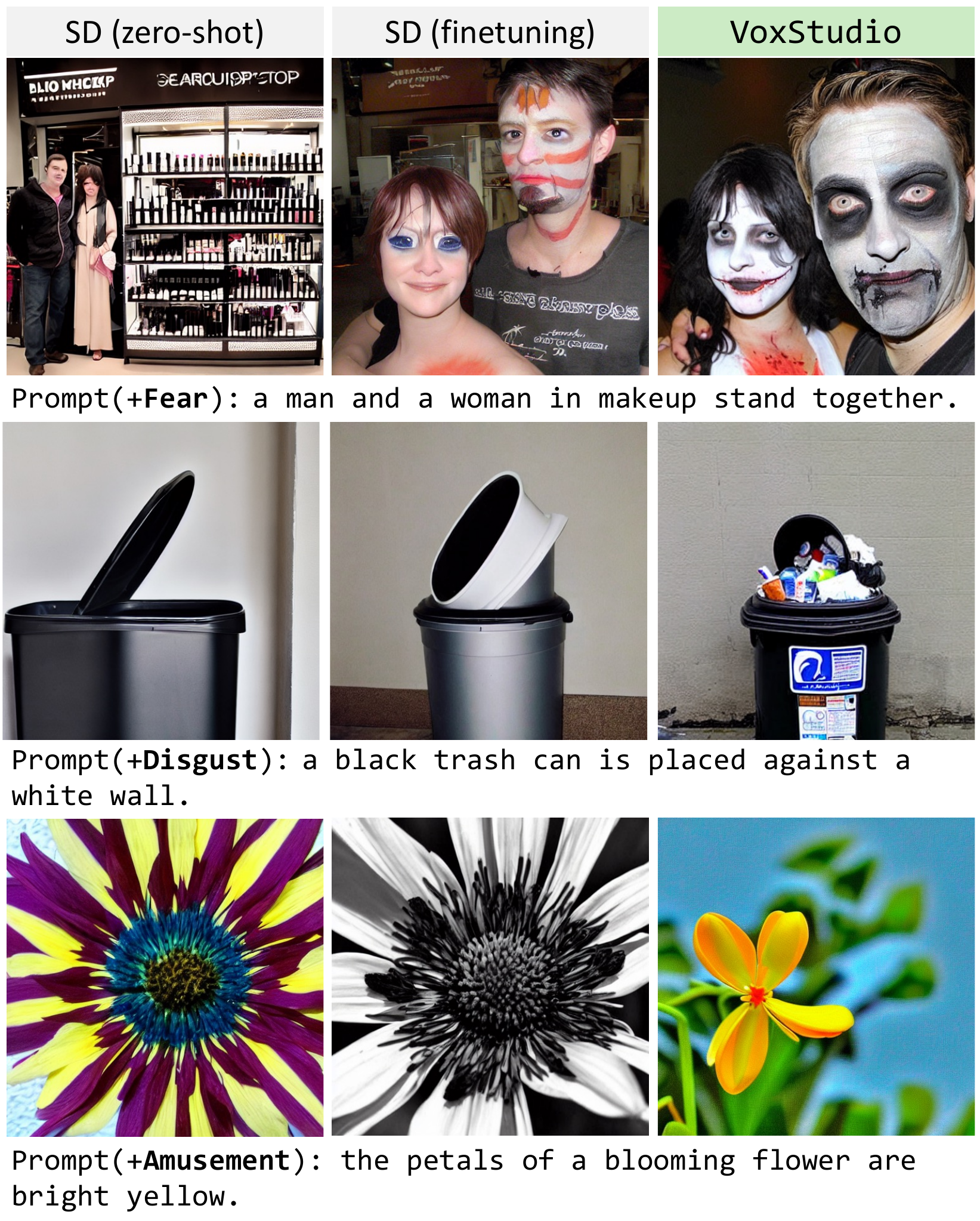}
    \vspace{-1.8em}
    \caption{{Qualitative comparison between SD using text prompts and \ours using speech prompts.}}
    \label{fig:sd_ours}
\end{figure}

%% file: table/spokenflickr.tex
\begin{table}[t]
    \centering
    \small
    \resizebox{0.8\linewidth}{!}{
    \begin{tabular}{lccc}
        \toprule
        Method & Zero-shot & FID$\downarrow$ & CLIPScore$\uparrow$  \\
        \midrule
        SpeechCLIP+ & \xmark & 63.19 & 23.71\\
        TMT & \xmark & 57.34 & 26.98 \\
        \ours & \cmark & \bf 55.01 & \bf 30.96  \\ 
        \bottomrule
    \end{tabular}
    }\vspace{-.5em}
    \caption{Performance comparison on Flikr8kAudio.}\label{tab:flickr}
\vspace{-.7em}        
\end{table}

%% file: fig_tex/scoco_flickr.tex
\begin{figure}[!t]
    \centering
    \begin{subfigure}[b]{\linewidth}
    \includegraphics[width=\linewidth]{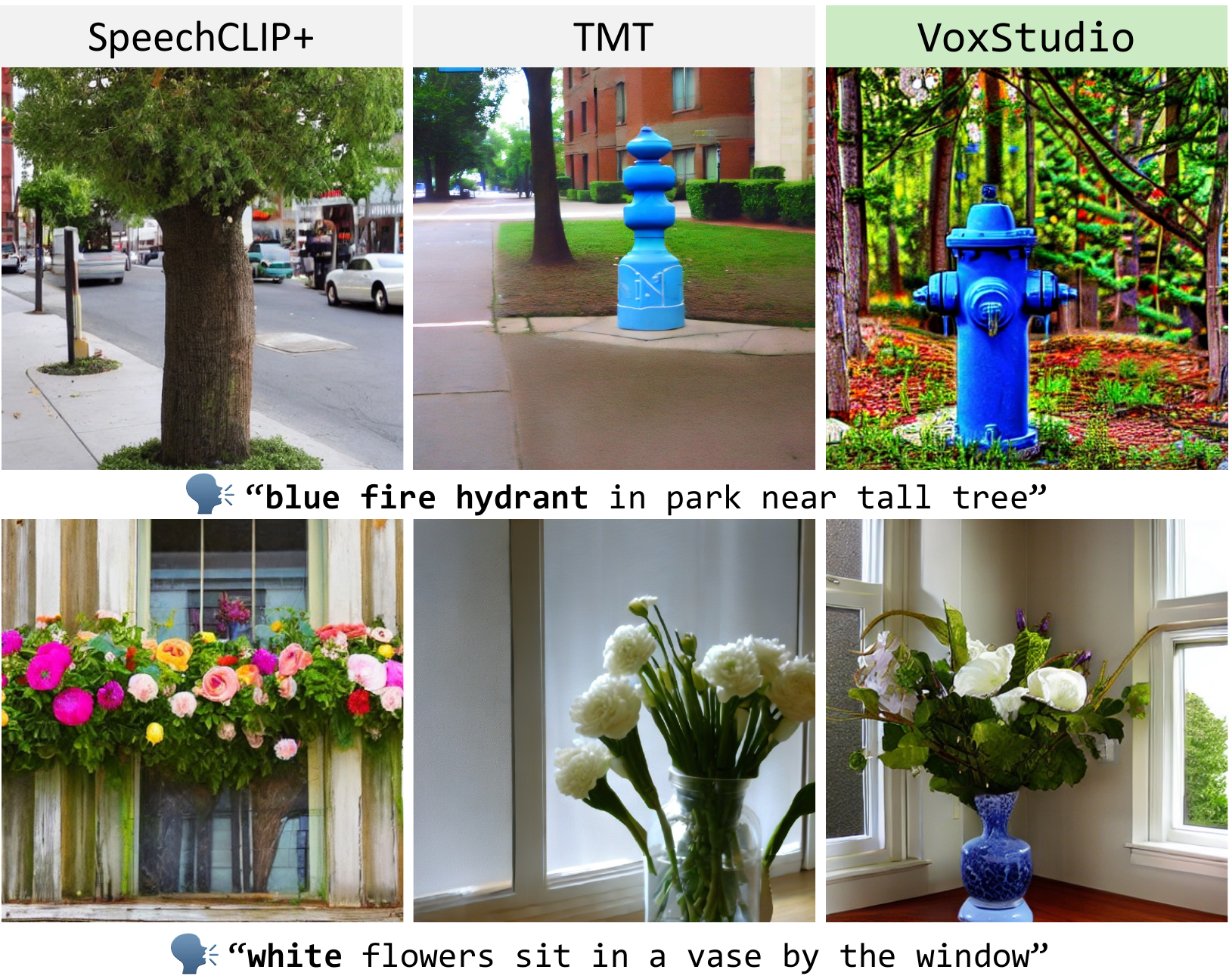}
    \caption{SpokenCOCO}\label{fig:a-scoco}
    \end{subfigure} \\
    \begin{subfigure}[b]{\linewidth}
    \includegraphics[width=\linewidth]{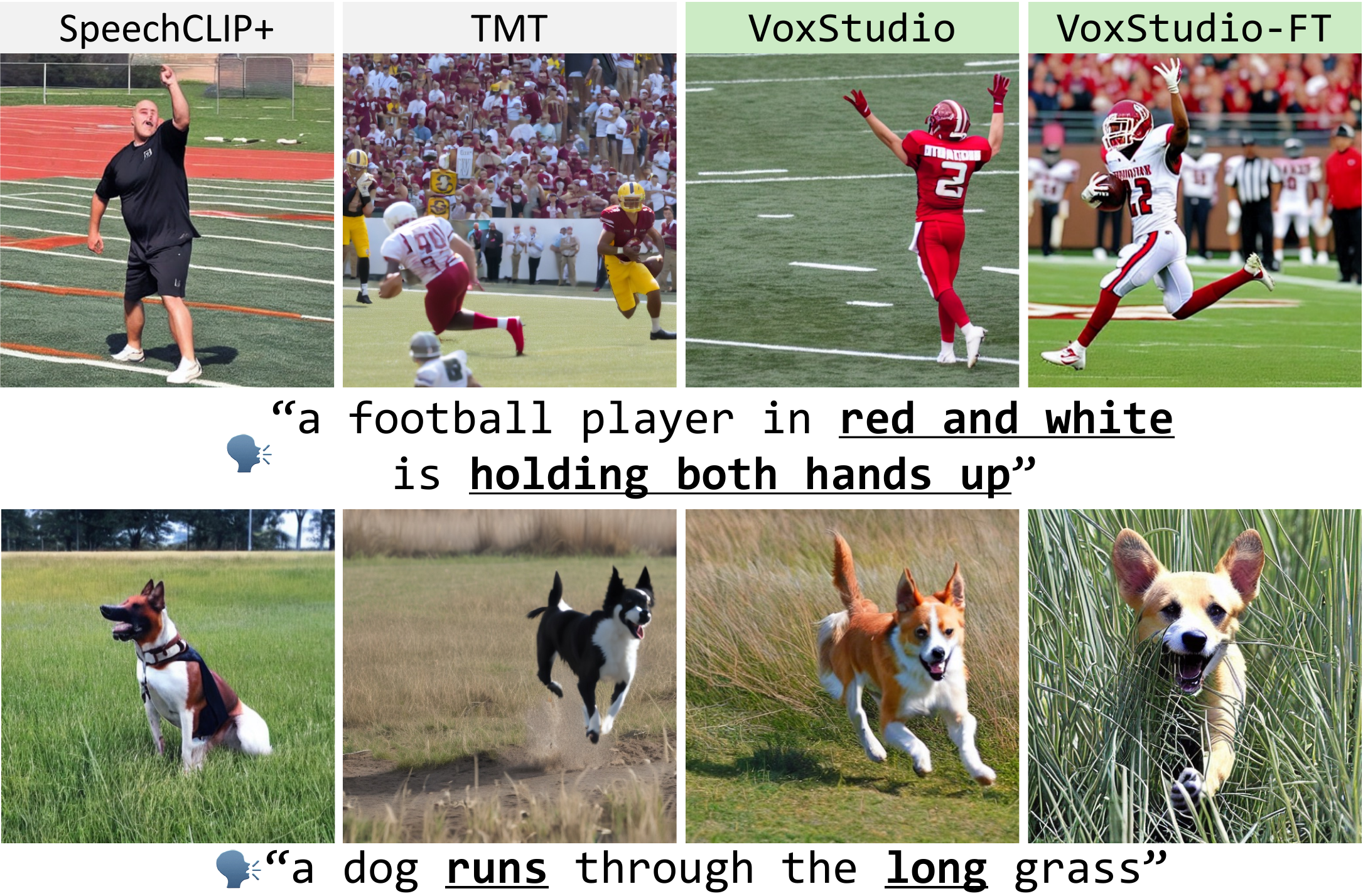}
    \caption{Flickr8kAudio}\label{fig:b-flickr}
    \end{subfigure}
    \vspace{-2em}
    \caption{{Qualitative comparisons on \scoco (first row) and Flickr8kAudio (second row). Compared to \ours, SpeechCLIP+ and TMT often miss out important concepts (\underline{underlined} words in examples).}}
    \label{fig:scoco_flickr}
\end{figure}

%% file: fig_tex/user_study.tex
\begin{figure}[t!]
    \centering
    \begin{subfigure}[b]{\linewidth}
    \centering
    \includegraphics[width=\linewidth]{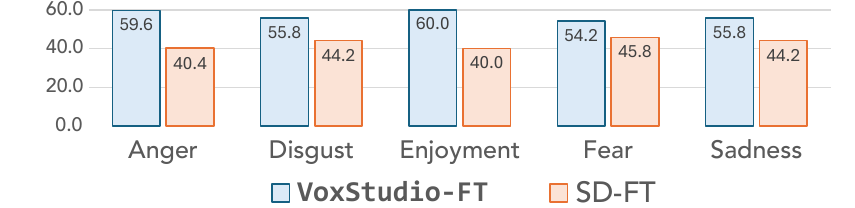}
    \caption{User study results on \oursbench}
    \label{fig:user_study1}
      \end{subfigure}
    \\
    \begin{subfigure}[b]{\linewidth}
    \centering
    \includegraphics[width=\linewidth]{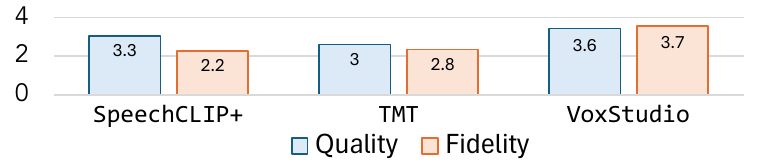}
    \caption{User study results on \scoco}
    \label{fig:user_study2}
      \end{subfigure}
      \vspace{-2em}
  \caption{{Human evaluation to evaluate (a) emotion consistency and (b) image quality and speech prompt fidelity.}}
    \label{fig:user_study}
    \vspace{-.5em}
\end{figure}

%% file: fig_tex/diff_emo.tex
\begin{figure}[t]
    \centering
    \includegraphics[width=.49\linewidth]{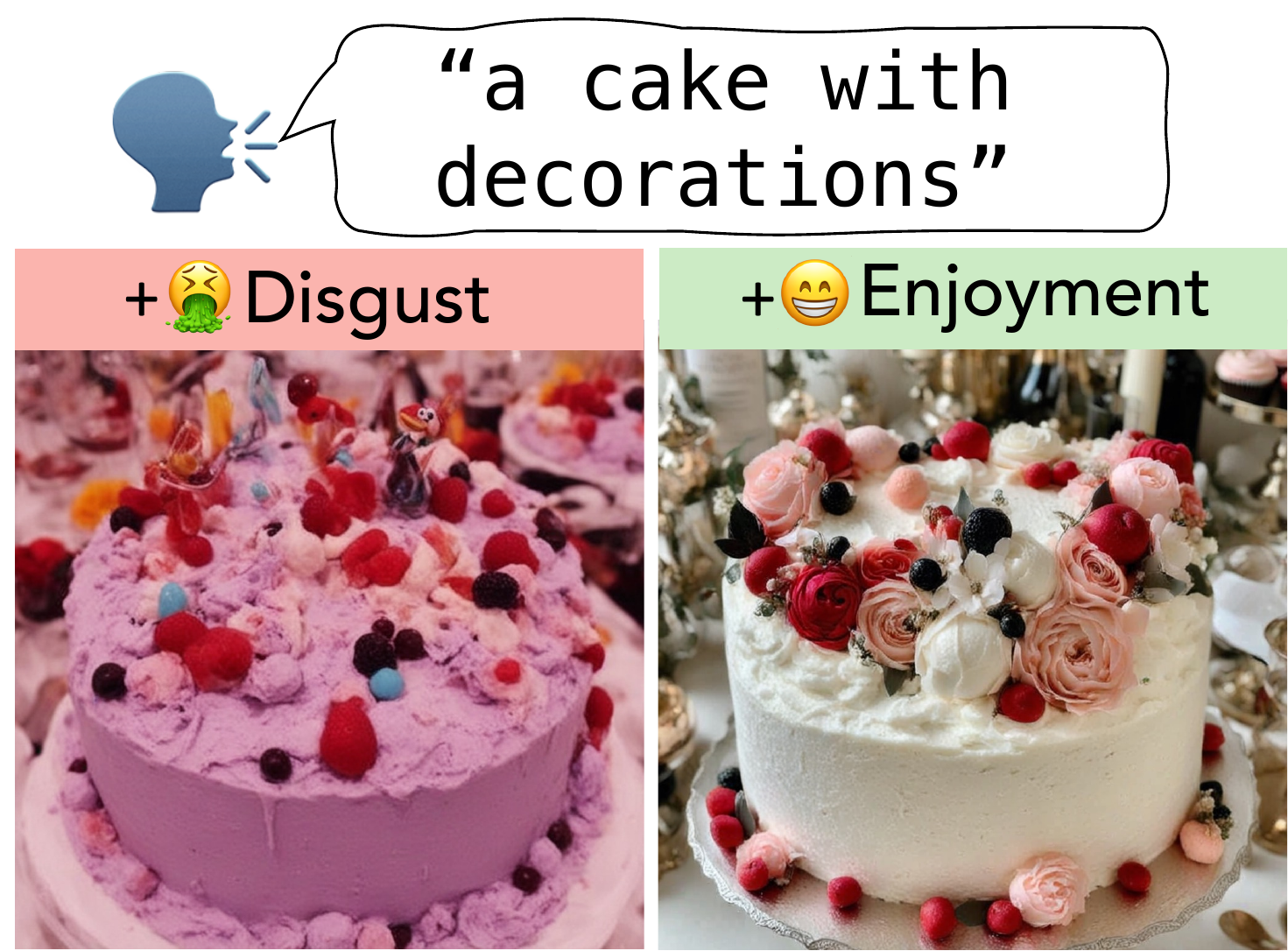}
    \includegraphics[width=.49\linewidth]{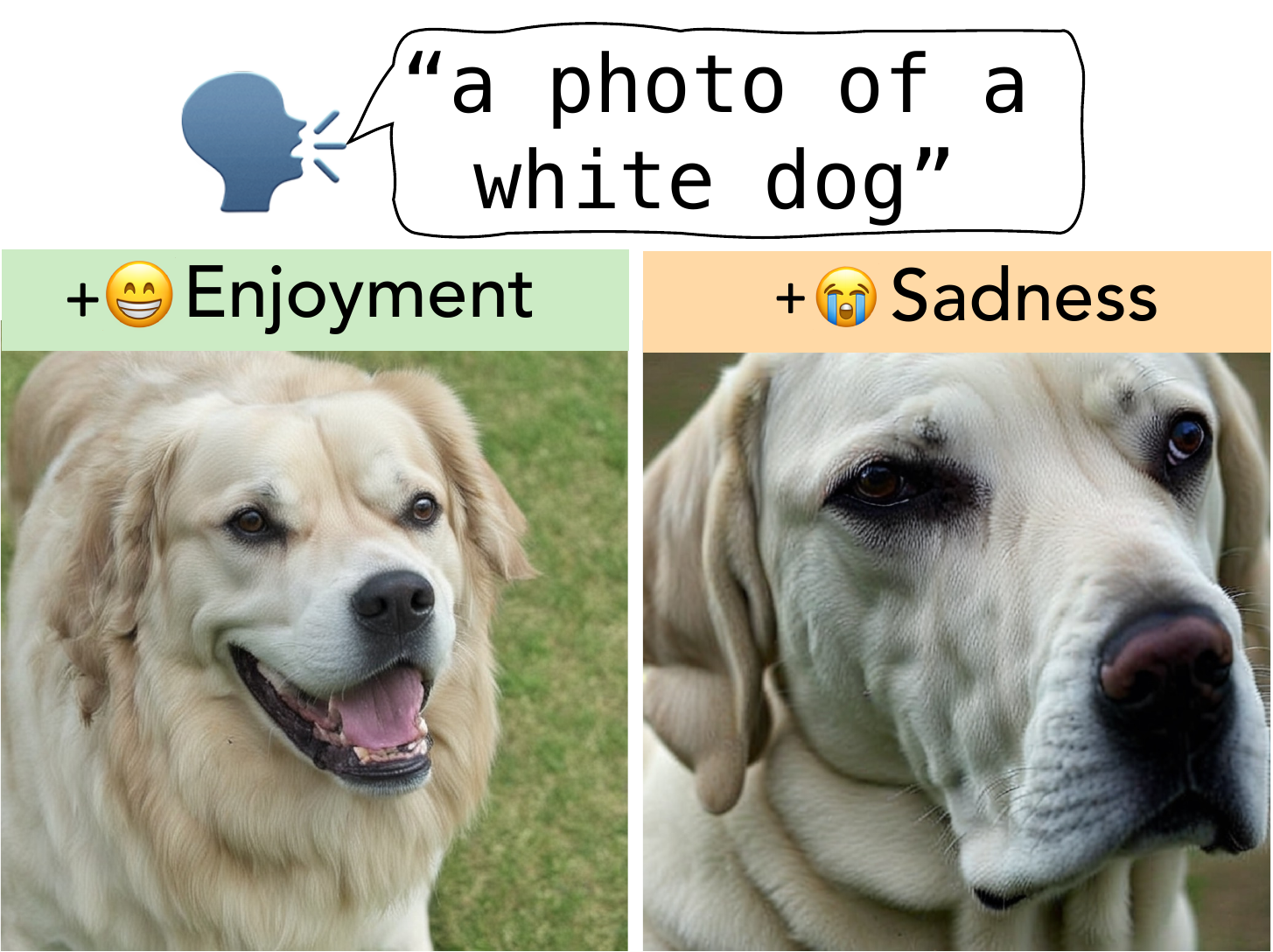}
    \vspace{-.7em}
    \caption{
    Generated images according to different emotions.
    Emotion in the voice evokes the sentimental changes in the generated image.
    }
    \label{fig:diff_emo}
\end{figure}

%% file: fig_tex/esd_real.tex
\begin{figure}[t!]
    \centering
    \includegraphics[width=\linewidth]{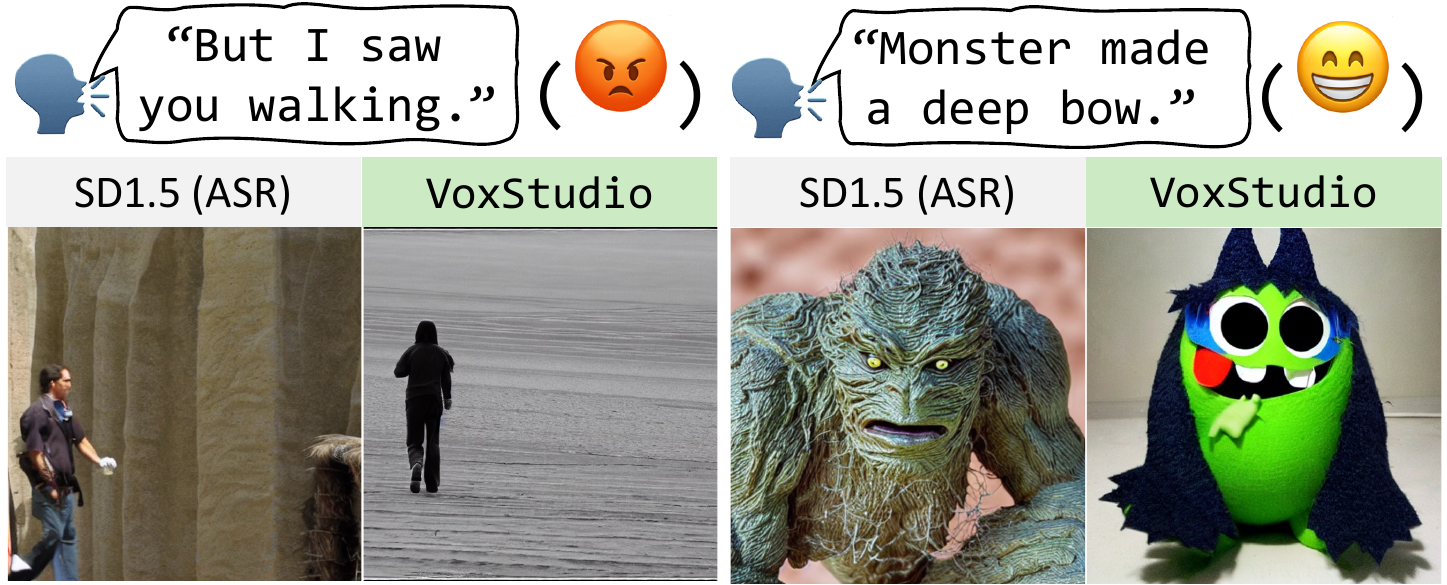}
    \vspace{-2em}
    \caption{Generated images from real humans' voice.}
    \label{fig:realgen}
\end{figure}

%% file: table/training_data.tex
\begin{table}[!t]
    
    \centering
    \small
    \resizebox{0.95\linewidth}{!}{
    \begin{tabular}{l ccc}
        \toprule
        Training data & FID$\downarrow$ & CLIPScore$\uparrow$ & Emo-A$\uparrow$ \\
        \midrule
        \scoco  &  32.60 & 26.16 & 46.20  \\
        \oursbench & 22.47 & 27.76 & 70.83  \\
        \scoco, \oursbench & \bf 19.94 & \bf 29.04 & \bf 71.70  \\
        \bottomrule
    \end{tabular}
    } 
    \vspace{-.5em}
    \caption{\small{Impact of the training datasets on \oursbench.}}\label{tab:training_data}
    \vspace{-1em}
\end{table}

%% file: table/finetuning.tex
\begin{table}[!t]  
    \centering
    \small
    \resizebox{\linewidth}{!}{
    \begin{tabular}{lccccc}
        \toprule
         Training & \# Tr. Params. & Input & FID$\downarrow$ & CLIPScore$\uparrow$ & Emo-A$\uparrow$ \\
        \midrule
        SD(T2I)-FT & 859.1M & T & \bf 18.31 & \bf 31.72 & 69.38 \\
        \midrule
        \ours &  50.0M & S & 25.01 & 28.71 & 67.09  \\
        \ours-LoRA &  50.7M & S & 27.25  & 29.88 & 69.43  \\
        \ours-FT &  909.1M & S & 19.94 & 29.04 & \bf 71.70  \\
        \bottomrule
    \end{tabular}
    } 
    \caption{Effect of the training strategies for SD1.5. We report the total number of trainable parameters.}\label{tab:finetuning}
\end{table}

%% file: table/sdxl_sd15.tex
\begin{table}[!t]
    \centering
    \small
    \resizebox{0.9\linewidth}{!}{
    \begin{tabular}{l c ccc}
        \toprule
        Base & UNet Size & FID$\downarrow$ & CLIPScore$\uparrow$ & Emo-A$\uparrow$ \\
        \midrule
        SD1.5 & 0.86B & 25.01 & \bf 28.71 & 67.09  \\
        SDXL & 2.6B &  \bf 23.12 & 28.04 &  \bf 69.26 \\
        \bottomrule
    \end{tabular}
    }
    \vspace{-.5em}
    \caption{{Effect of the scale of image generator.}}\label{tab:sdxl}
\end{table}

%% file: table/encoder.tex
\begin{table}[t]
    \centering
    \small
    \resizebox{0.97\linewidth}{!}{
        \begin{tabular}{lcccc}
            \toprule
            Encoder & \# Params. & FID$\downarrow$ & CLIPScore$\uparrow$ & Emo-A$\uparrow$ \\
            \midrule
            Whisper-L v3 &  636M & 23.57 & 28.33 & 67.77  \\
            SONAR &  600M & \bf 19.94 & \bf 29.04 & \bf 71.70  \\
            \bottomrule
        \end{tabular}
    }\vspace{-.5em}
    \caption{Impact of the encoder choices.}\label{tab:encoder}
\end{table}

%% file: table/mapper.tex
\begin{table}[!t]
    \centering
    \small
    \resizebox{0.97\linewidth}{!}{
    \begin{tabular}{lcccc}
        \toprule
        SIB architecture & \# Params. & FID$\downarrow$ & CLIPScore$\uparrow$ & Emo-A$\uparrow$ \\
        \midrule
        Transformer &  71M & 23.12 & 28.04 & 69.26  \\
        \ours & 50M & \bf 19.94 & \bf 29.04 & \bf 71.70  \\
        \bottomrule
    \end{tabular}
    }
    \vspace{-.5em}
    \caption{{Effectivness of architecture choices of \mapper.}}\label{tab:mapper}
    \vspace{-1em}
\end{table}

%% file: tex/6_conclusion.tex
\section{Conclusion}
\ours is the first end-to-end S2I model that captures both linguistic and emotional nuances from speech.
Unlike text-based methods, our approach totally leverages speech’s expressiveness to generate emotionally aligned images. \oursbench is built cheaply, but it is complementary with real-world datasets.
Our experiments demonstrate that \ours not only outperforms prior speech-based methods in conveying sentiment through images, but also matches text-driven approaches in semantic alignment, despite the higher noise and lower latency of the speech modality.
We believe our work facilitates future research in voice-driven generative models and their applications.

%% file: supp_tex/a_benchmark.tex
\section{\oursbench}\label{app:benchmark}
Our objective for data construction is primarily on (1) synthesizing large-scale image and speech pairs, (2) the speech will be emotional rich, and (3) closely matches the quality of real recordings while diversifying the range of speakers.
Here we supplement the details of \oursbench.

First, we collect the images in the 118k subset of EmoSet~\cite{emoset}.
To balance the positive and negative emotions, we consolidate `amusement' and `excitement' into the `enjoyment' class.
We use the original train and test split.

We use the following prompt to generate text captions for images using LLaVA-OneVision~\cite{llava-onevision}:

\noindent
\begin{tabularx}{\linewidth}{X}
\toprule
    \textit{Generate three disjoint captions for the given image. Each caption should:} \\
    \textit{* Have a different sentence structure,} \\ 
    \textit{* Avoid emotional or subjective expressions,} \\
    \textit{* Describe different aspects of the image, such as objects, actions, spatial relationships, or surroundings,} \\
    \textit{* Be between 8 and 15 words long,} \\
\bottomrule \vspace{1pt}
\end{tabularx}
\noindent
Following the characteristics of \scoco, we limit the length of captions and ensure they accurately describe the context of the image.
In \cref{app:word_dist}, the word count distribution remains similar to \scoco and \flickr, but with a more structured and consistent pattern.

For speech generation, we use state-of-the-art F5-TTS~\cite{f5tts}, where the vocoder is trained on 24kHz. 
The generated speech is resampled to 16kHz to use SONAR and Whisper encoders.
To diversify the speaker characteristics, we collect multiple emotional speech datasets: MEAD~\cite{mead}, CREMA-D~\cite{crema-d}, and RAVDESS~\cite{ravdess}, which are widely used in emotional speech synthesis and speech emotion recognition.
MEAD is an audiovisual dataset annotated in 8 emotional categories. 
CREMA-D is a crowd-sourced actor dataset, using 6 emotion classes.
RAVDESS contains audio and video, where the professional actors express emotion.
All datasets used English.
The split cleaned up by EmoBox~\cite{emobox} is used, especially the fold 1 split for RAVDESS.
\cref{app:emodataset} summarizes the number of emotion classes and speakers for each dataset.

%% file: supp_tex/b_result.tex
\section{More Results}

\subsection{Ablation study}
The results for \ours on \scoco according to the difference of training data is shown in \cref{tab:training_data_coco}.
In the results, we observe that \scoco does not fully capture the diversity of real-world scenarios. 
Most images convey neutral emotions and primarily depict scenes suitable for objective descriptions. 
However, real-world photographs go beyond presenting mere facts, often communicating higher-level meanings such as feelings~\cite{yang2024emogen}. 
This demonstrates that our dataset extends beyond the distribution of \scoco and \flickr, offering a more realistic and emotionally expressive representation.
This claim is further supported by Table 2 of the main paper, where models trained on the \scoco and \flickr datasets— even TMT, which was trained on a massive amount of synthesized speech from CC3M and CC12M—fail to generalize to our dataset.

\begin{figure}[t]
    \centering
    \includegraphics[width=.95\linewidth]{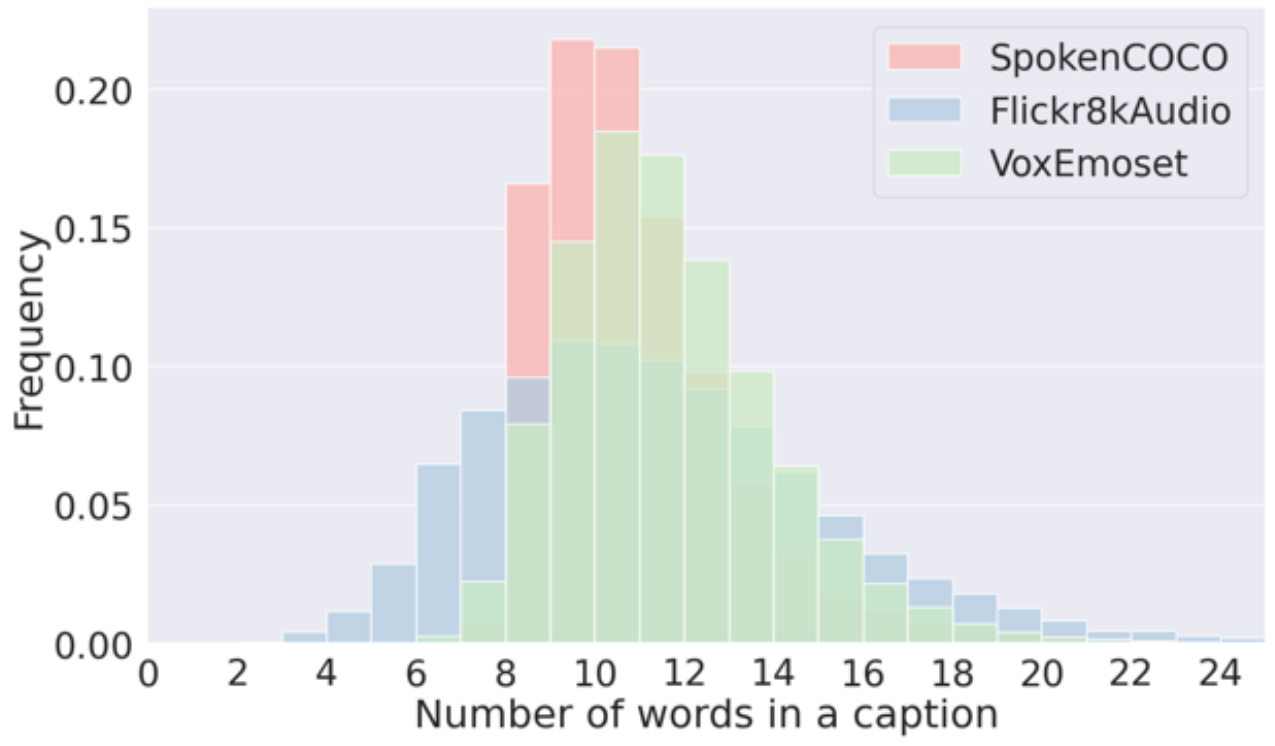}
    \vspace{-1em}
    \caption{Histogram of the number of words in each description in SpokenCOCO~\cite{scoco}, Flickr8kAudio~\cite{spokenflickr} and \oursbench (ours).}
    \label{app:word_dist}
\end{figure}

\input{table/emotion_data}

\input{table/training_data_coco}

\input{fig_tex/app_sdft}
\input{fig_tex/app_sdall}

\subsection{Qualitative results}
We show more qualitative comparisons in \cref{fig:app_sdft} and \cref{fig:app_sdall}.
In those experiments, our image generators are initialized by UNet parameters in SD1.5.
When training SD with a CLIP encoder using text prompts from our dataset, the model better follows the content of the text compared to zero-shot generation. 
However, in terms of emotional intensity and expressiveness, it performs weaker than our approach using speech prompts.
For example, in the last row in \cref{fig:app_sdall}, generating an emotionally rich image from a sentence like `a tomato is cut into sections on a white plate' is challenging. 
By using speech input that conveys a feeling of disgust, our model generates an image where the tomato appears distorted, conditioned on the given emotion category.
Moreover, despite the inherent noise in speech, \ours utilizes SIB to refine the information and capture its meaning, effectively following the content of the prompt.

\cref{fig:app_ours_more} and \cref{fig:app_sdxl} provide more results generated using the parameters of SD1.5 and SDXL, respectively.
Especially for the example of `a woman with blonde hair is standing in a room.', \ours poses the sadness in her facial expression.
While we freeze parameters of the image generator, \cref{fig:app_sdxl} shows that generated images ensure high fidelity and text relevancy.
\input{fig_tex/app_ours_more}
\input{fig_tex/app_sdxl}

\subsection{Details of human evaluation}
To assess how well our model captures paralinguistic information in speech, we conducted a human evaluation to measure the alignment between the emotions perceived in the input speech and those conveyed in the generated images.
We had 26 locally recruited participants evaluate 25 images.
Five images represent each emotion, but we did not provide any information about which emotion each speech sample conveyed. 
Participants evaluated the images with randomly mixed emotion classes.
The instruction in \cref{app:user_inst} is used in human evaluation.

We recruit 17 independent evaluators to assess image quality and speech prompt fidelity on \scoco in Fig. 7(b) of the main paper. 
In this experiment, the instructions in \cref{app:user_inst2} are used for evaluating 10 different images generated by SpeechCLIP+, TMT, and \ours, respectively.
\input{fig_tex/app_inst}

\input{fig_tex/app_inst2_quality}

\begin{figure}[t]
    \centering
    \includegraphics[width=\linewidth]{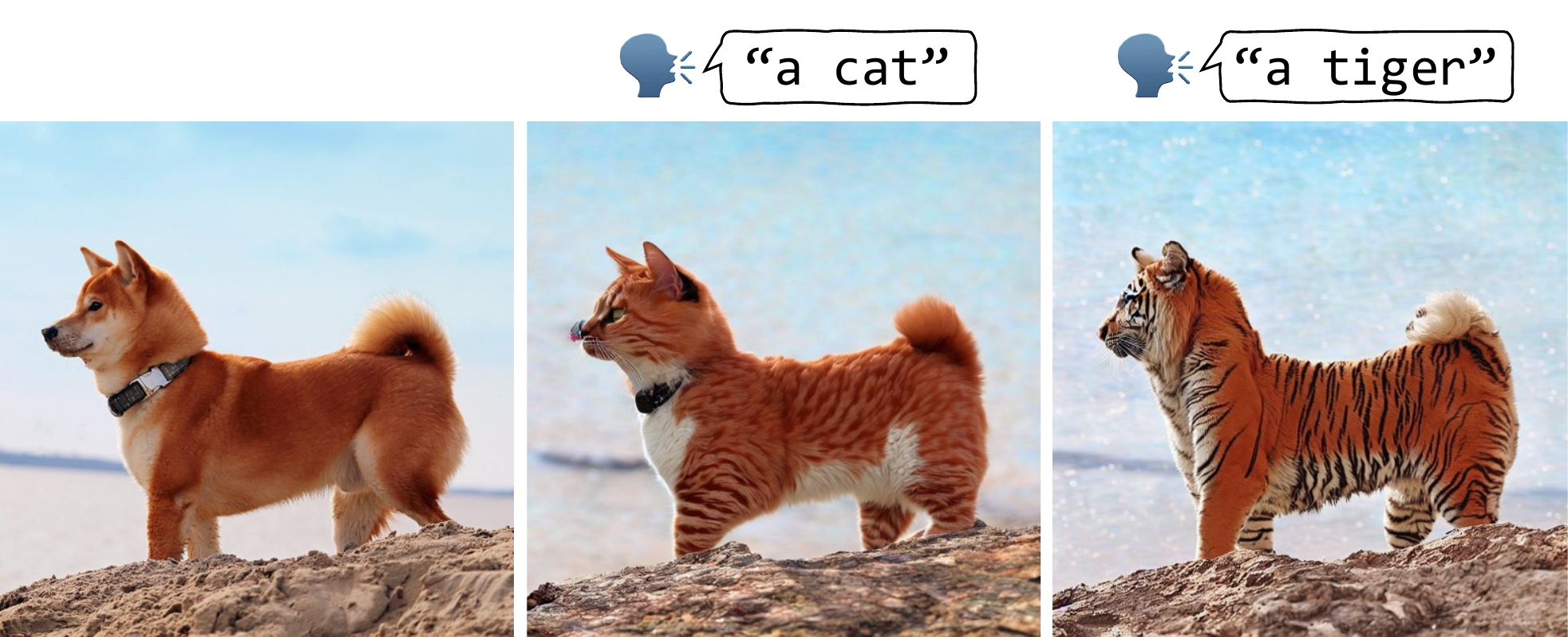}
    \caption{Image editing using speech prompt.}
    \label{fig:i2i}
\end{figure}

\subsection{Failure cases and limitation}\label{app:limit}
Although it is impossible to manually verify all samples, we found that SpokenCOCO dataset, which was created using human annotators via AMT, often contains misrecorded speech samples. 
For example, as shown in \cref{fig:a-limit}, some recordings mispronounce the text prompt originally associated with the image.
Therefore, using speech inputs that contain such errors is inevitably prone to performance degradation compared to using text inputs.
Additionally, the clarity of word representation in speech depends on the speaker’s pronunciation, making it challenging to distinguish homophones or similarly pronounced words.

\input{fig_tex/limitation}

%% file: table/emotion_data.tex
\begin{table}[t]
    
    \centering
    \small
    \begin{tabular}{lcc}
        \toprule
        Dataset & Classes & \# Speakers \\
        \midrule
        RAVDESS & 8 & 24 \\
        MEAD & 8 & 48\\
        CREMA-D & 6 & 91 \\
        \bottomrule
    \end{tabular}

    \caption{Characteristics of speech emotion datasets used as emotion and speaker condition.}\label{app:emodataset}

\end{table}

%% file: table/training_data_coco.tex
\begin{table}[t]
    \centering
    \small
    \begin{tabular}{cc cc}
        \toprule
        \scoco & \oursbench & FID$\downarrow$ & CLIPScore$\uparrow$ \\
        \midrule
        \cmark & \xmark & \bf 24.95 & \bf 29.04  \\
        \xmark & \cmark & 32.59 & 25.32  \\
        \cmark & \cmark & 27.15 & 27.27  \\
        \bottomrule
    \end{tabular}

    \caption{Impact of the training datasets on \scoco.}\label{tab:training_data_coco}
\end{table}

%% file: fig_tex/app_sdft.tex
\begin{figure*}[ht!]
    \centering
    \includegraphics[width=0.9\linewidth]{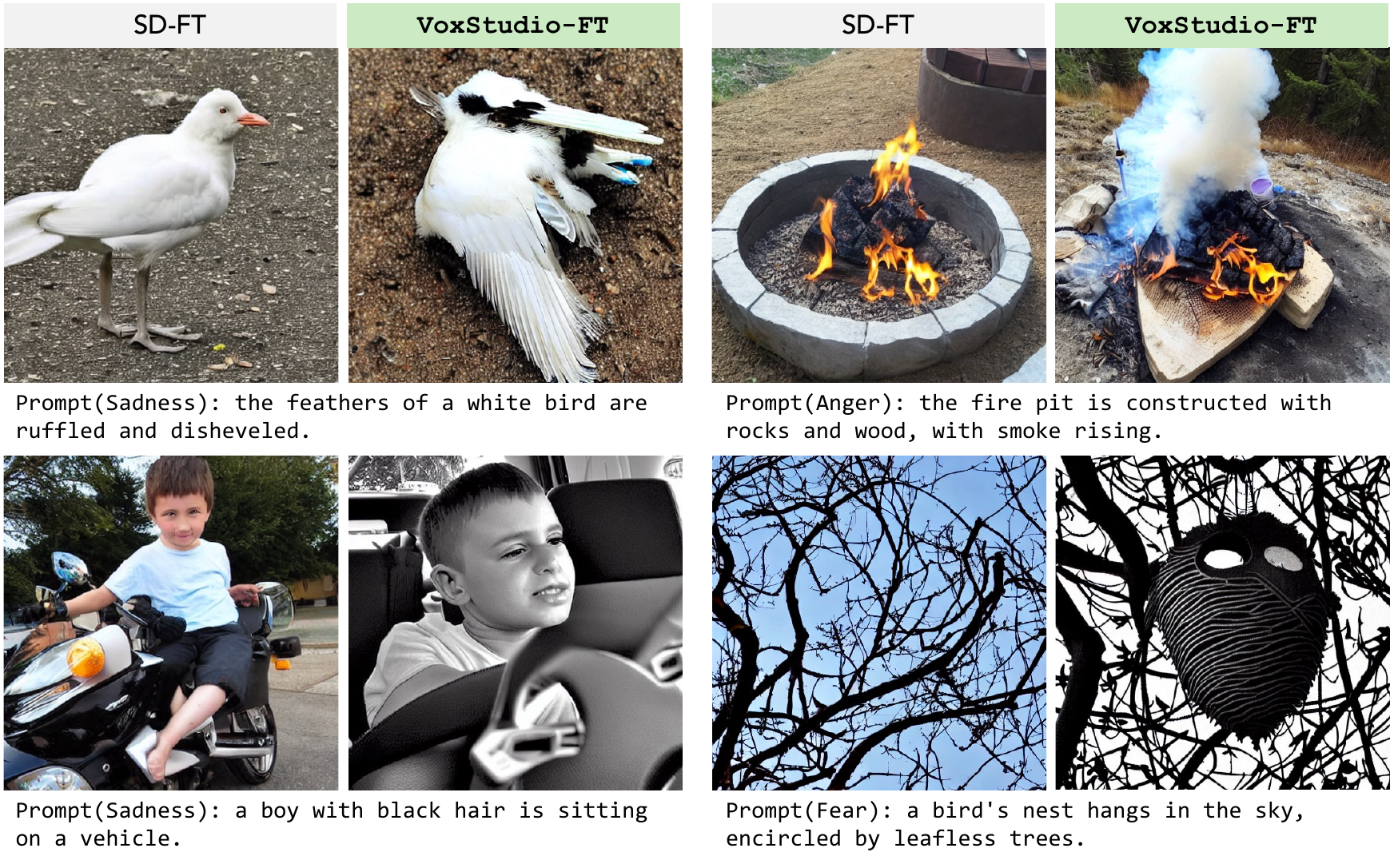}
    \vspace{-10pt}
    \caption{Qualitative comparison between SD1.5 finetuned with text prompts and \ours-FT trained with speech prompts.}
    \label{fig:app_sdft}
\end{figure*}

%% file: fig_tex/app_sdall.tex
\begin{figure}[t!]
    \centering
    \includegraphics[width=\linewidth]{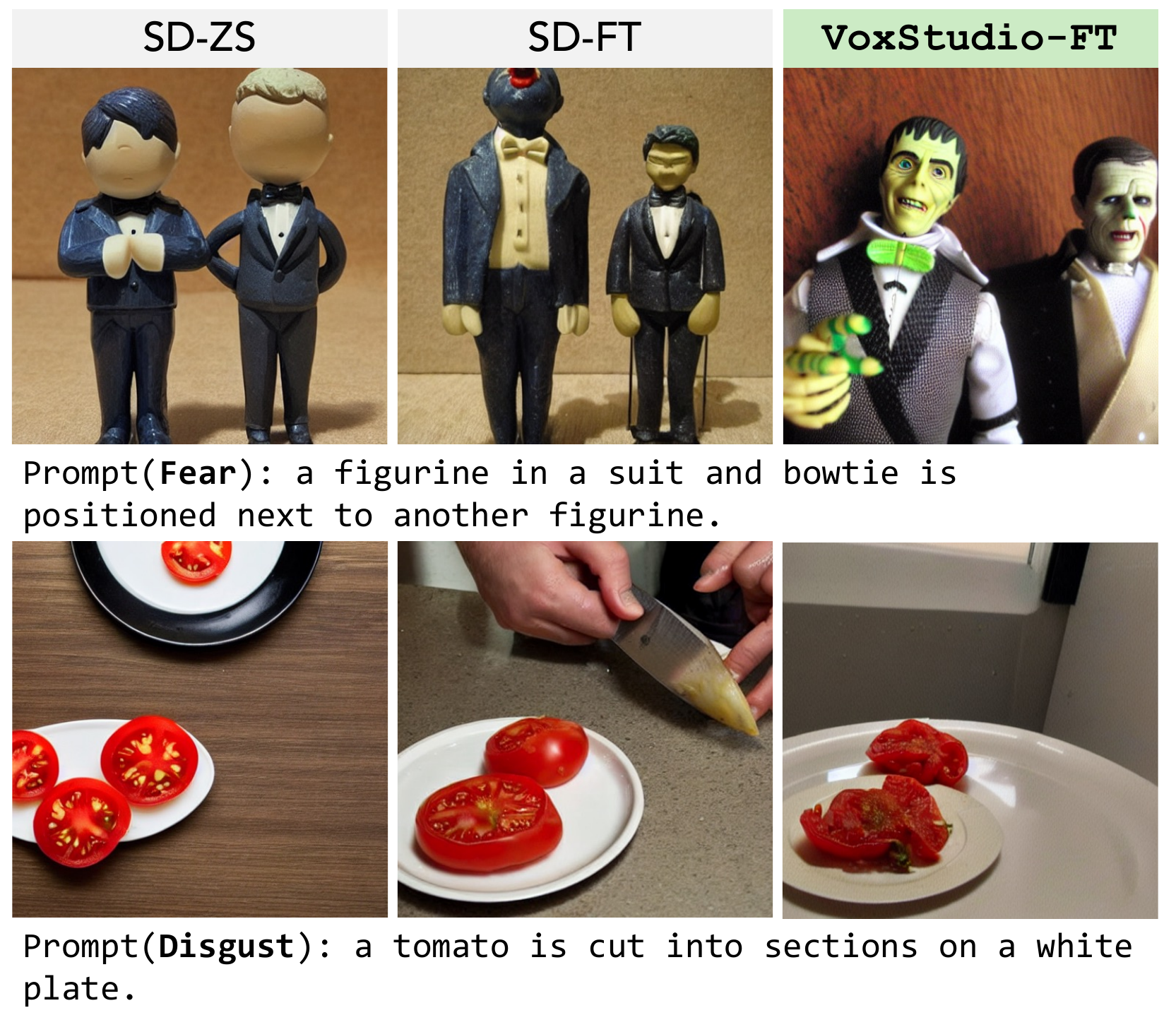}
    \vspace{-20pt}
    \caption{Qualitative comparison between SD1.5 zero-shot generation \vs SD1.5 finetuned with text prompts \vs \ours-FT trained with speech prompts.}
    \label{fig:app_sdall}
\end{figure}

%% file: fig_tex/app_ours_more.tex
\begin{figure*}[ht!]
    \centering
    \includegraphics[width=\linewidth]{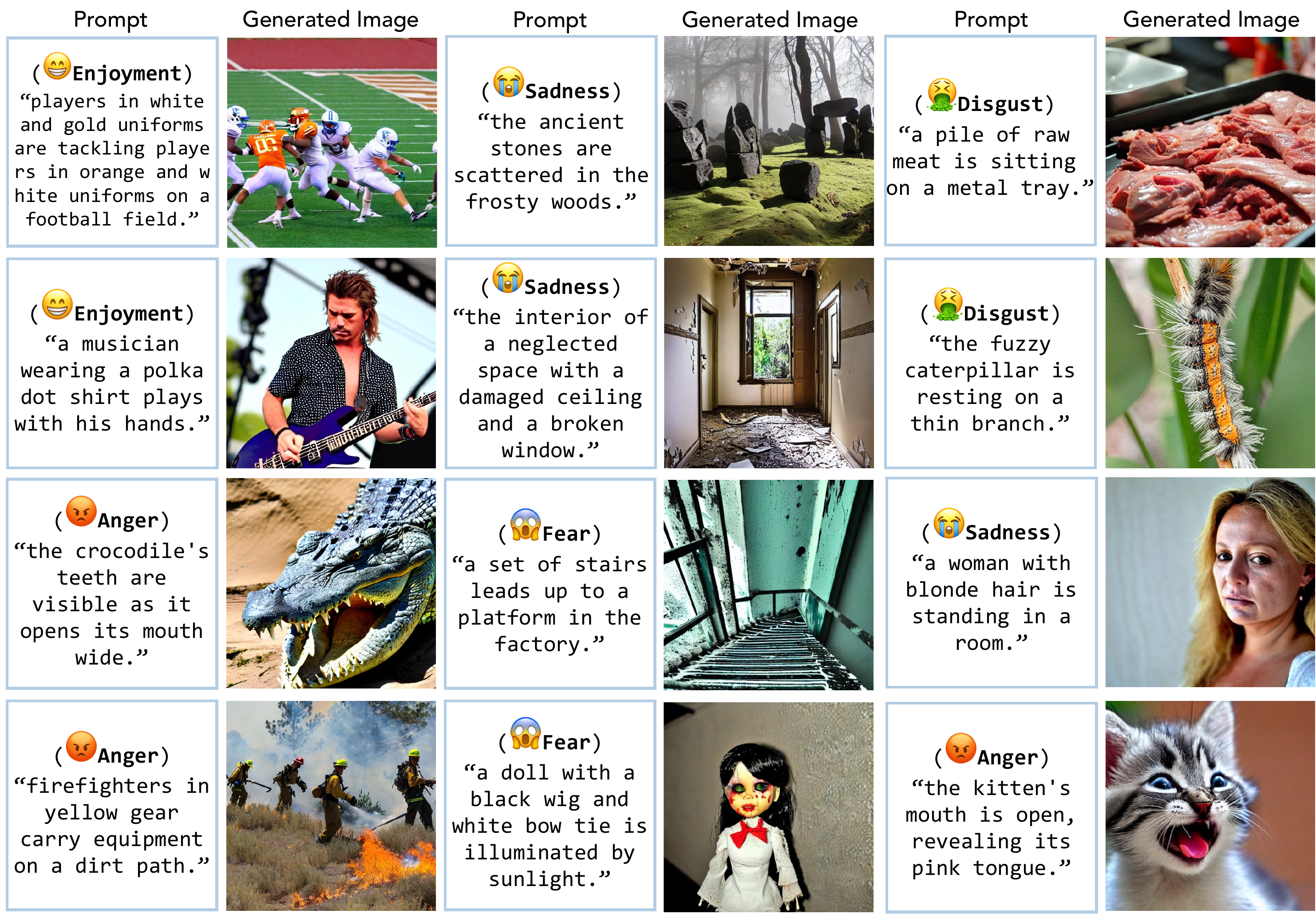}
    \caption{More qualitative examples generated by \ours-FT.}
    \label{fig:app_ours_more}
\end{figure*}

%% file: fig_tex/app_sdxl.tex
\begin{figure*}[ht!]
    \centering
    \includegraphics[width=\linewidth]{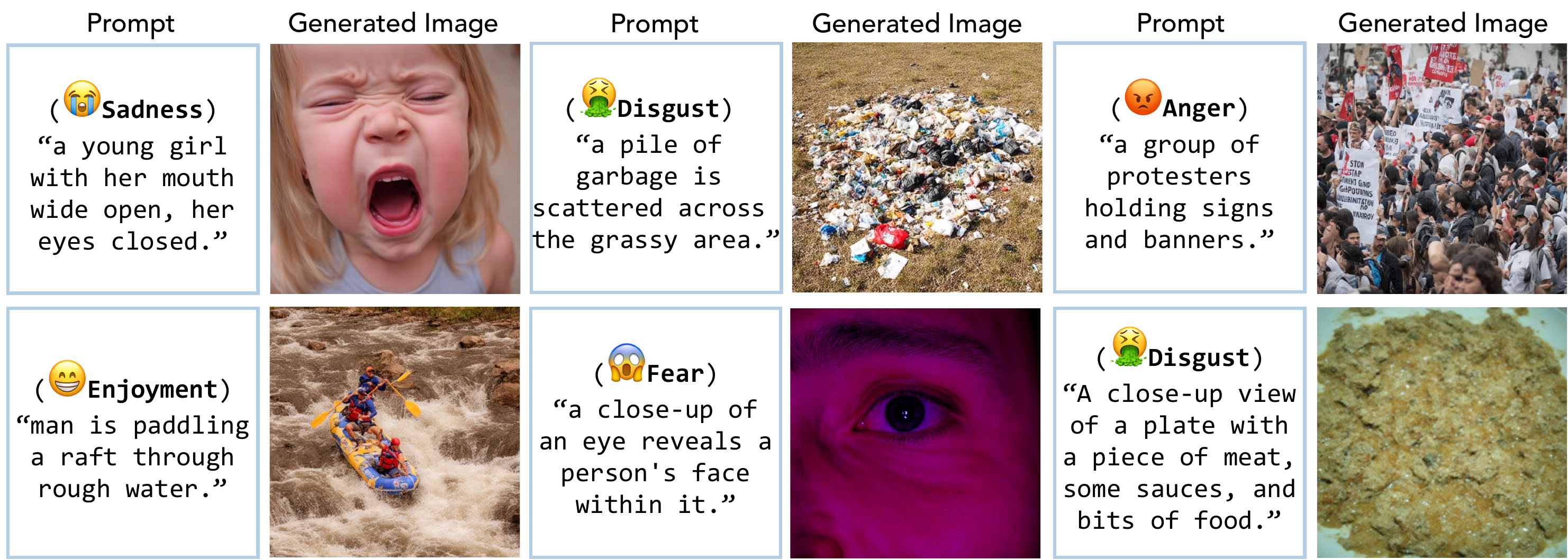}
    \caption{Qualitative results of \ours, where the parameters of image generator from SDXL. We freeze the image generator during the training.}
    \label{fig:app_sdxl}
\end{figure*}

%% file: fig_tex/app_inst.tex
\newtcbtheorem{Instruct}{\bfseries Instruction}{enhanced,drop shadow={black!50!white},
  coltitle=black,
  top=0.3in,
  attach boxed title to top left=
  {xshift=1.5em,yshift=-\tcboxedtitleheight/2},
  boxed title style={size=small,colback=yellow}
}{instruction}

\newtcolorbox[auto counter]{instruction}[1][]{title={\bfseries Instruction},enhanced,drop shadow={black!50!white},
  coltitle=black,
  top=0.3in,
  attach boxed title to top left=
  {xshift=1.5em,yshift=-\tcboxedtitleheight/2},
  boxed title style={size=small,colback=yellow},}

\begin{figure}[t]
    \centering
    \begin{Instruct}{}{first}
    Preference test: \\
    Which image better represents the given speech in terms of emotional expression (one of the enjoyment, fear, disgust, anger, and sadness)?
    \\
    \\
    Instructions: \\
    1. Click the play button to listen to the speech clip. \\
    2. Observe the two images displayed below. \\
    3. Choose which image you think better represents the emotion in speech. \\
        \vspace{0.5em}
    \end{Instruct}
\caption{User instruction used in human evaluation.}\label{app:user_inst}
\end{figure}

%% file: fig_tex/app_inst2_quality.tex
\begin{figure}[t]
    \centering
    \begin{Instruct}{}{first}

    1. Image Quality Test:\\\\
Please rate the realism of the generated image. 
Do not consider how well it matches the prompt.\\

1 = Extremely unrealistic\\
2 = Somewhat unrealistic\\
3 = Neither realistic nor unrealistic \\
4 = Somewhat realistic \\
5 = Extremely realistic \\

2. Prompt Fidelity Test:\\\\
Listen to the spoken prompt and rate how accurately the image reflects its content. 
Evaluate details such as objects, shapes, backgrounds, and other elements. 
Do not consider the image’s overall quality.\\

1 = Not at all consistent\\
2 = Partially consistent\\
3 = Moderately consistent\\
4 = Very consistent\\
5 = Perfectly consistent\\

        \vspace{0.5em}
    \end{Instruct}
\caption{User instructions used in human evaluation.}\label{app:user_inst2}
\end{figure}

%% file: fig_tex/limitation.tex
\begin{figure}[!t]
    \centering
    \begin{subfigure}[b]{\linewidth}
    \includegraphics[width=\linewidth]{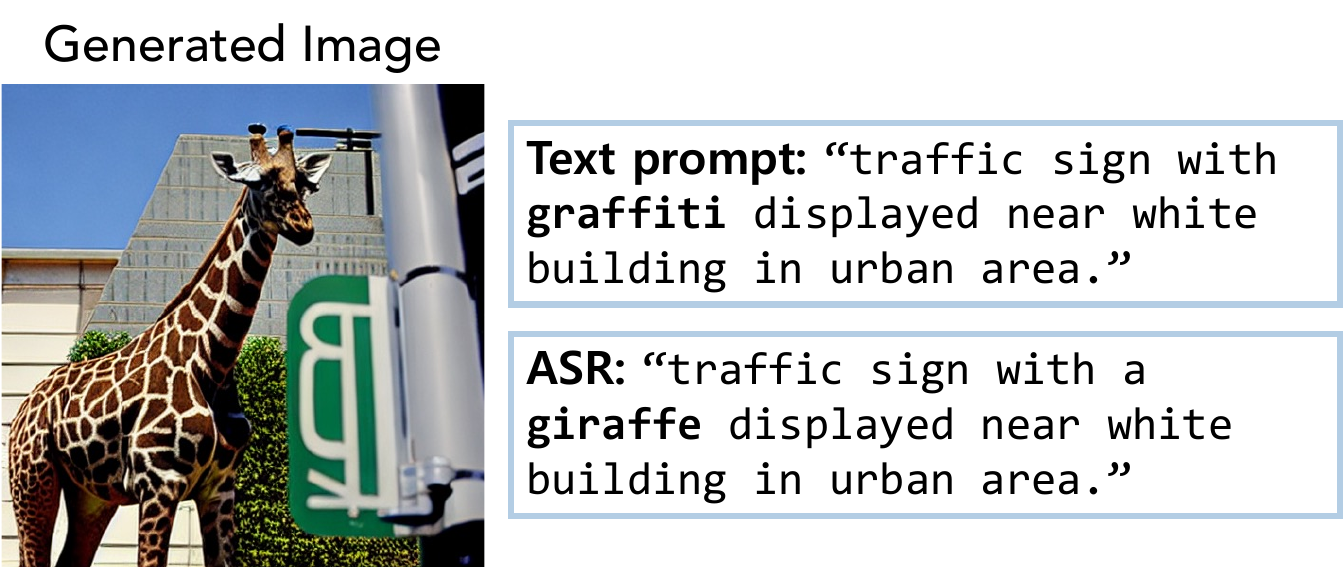}
    \caption{}\label{fig:a-limit}
    \end{subfigure} \\
    \begin{subfigure}[b]{\linewidth}
    \includegraphics[width=\linewidth]{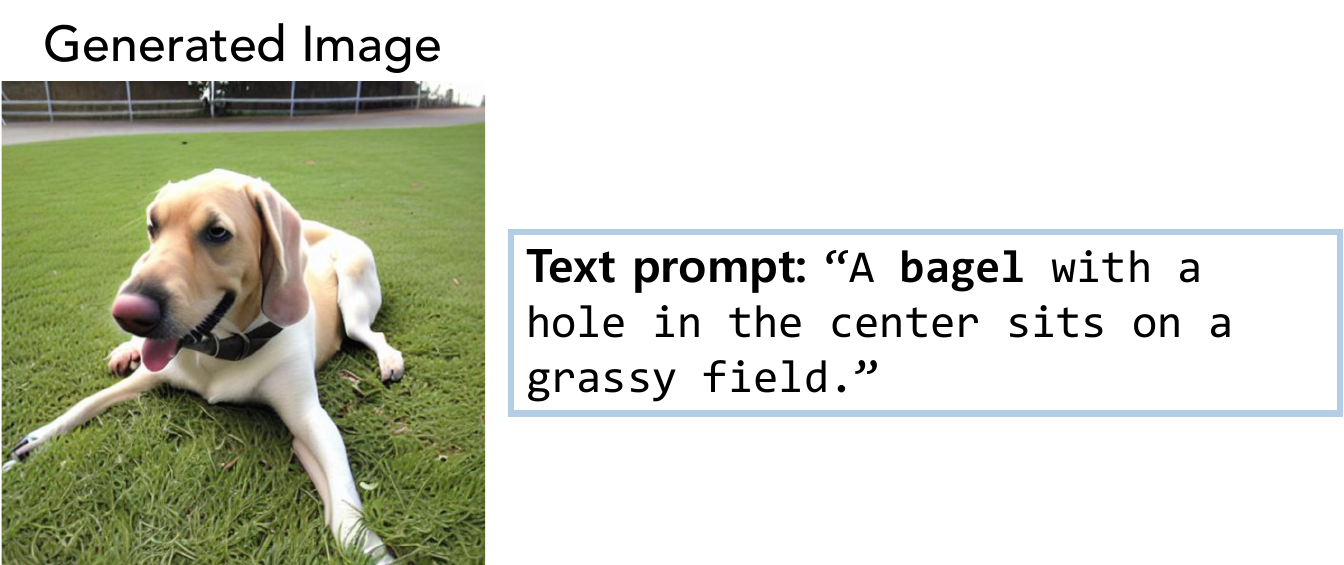}
    \caption{}\label{fig:b-limit}
    \end{subfigure}
    \caption{Failure cases in (a) misreading words (graffiti vs. giraffe) in \scoco, and (b) confusion between words with similar pronunciation (bagel vs. Beagle).}
    \label{fig:app_limit}
\end{figure}